\documentclass[letterpaper,11pt]{article}
\usepackage{jheppub}
\pdfoutput=1
\usepackage{amsmath,amssymb,amsfonts,bm,amscd,mathtools}
\usepackage{graphicx}
\usepackage{multirow}
\usepackage{verbatim}
\usepackage{appendix}
\usepackage{fancybox}
\usepackage{array}
\usepackage{url}
\usepackage{float}
\usepackage{subfigure}

\def\CN{{\mathcal N}}
\def\CO{{\mathcal O}}

\def\be{\begin{equation}}
\def\ee{\end{equation}}
\def\bea{\begin{eqnarray}}
\def\eea{\end{eqnarray}}

\newcommand{\RR}{{\mathbb R}}
\newcommand{\eW}{{\cal W}}
\newcommand{\eX}{{\cal X}}
\newcommand{\eY}{{\cal Y}}
\newcommand{\eZ}{{\cal Z}}
\newcommand{\eP}{{\cal P}}
\newcommand{\eQ}{{\cal Q}}
\newcommand{\eC}{{\cal C}}
\newcommand{\eT}{{\cal T}}
\newcommand{\pW}{{W}}
\newcommand{\pX}{{X}}
\newcommand{\pY}{{Y}}
\newcommand{\pZ}{{Z}}
\newcommand{\pP}{{P}}

\newcommand{\pL}{{L}}
\newcommand{\tR}{{\mathtt R}}

\newcommand{\sq}{{\mathsf q}}
\newcommand{\sfp}{{\mathsf p}}

\newcommand{\fS}{{\mathsf S}}

\title{A Scattering Amplitude for Massive Particles in AdS}

\author{Abhijit Gadde and Trakshu Sharma}
\affiliation{Department of Theoretical Physics \\ 
Tata Institute for Fundamental Research, Mumbai 400005}
\emailAdd{abhijit@theory.tifr.res.in, trakshu.sharma@tifr.res.in}

\abstract{
In this paper, we propose a conformally covariant momentum space representation of CFT correlation functions. We call it the AdS S-matrix. This representation has the property that it reduces to the S-matrix in the flat space limit. The flat space limit in question is taken by keeping all the particle masses fixed as the operator conformal dimensions go to infinity along with the AdS radius $1/ \mathtt{R}$. We give Feynman-like rules to compute the AdS S-matrix in $1/ \mathtt{R}$ perturbation theory. Moreover, we relate it to the Mellin space representation of the conformal correlators in $1/ \mathtt{R}$ perturbation theory.}

\begin{document}
\maketitle
\flushbottom

\section{Introduction}
What is the local observable in the bulk theory that the conformal correlator computes? In the so called flat space limit \cite{Polchinski:1999ry, Giddings:1999jq, Gary:2009ae, Okuda:2010ym, Penedones:2010ue,  Maldacena:2015iua, Paulos:2016fap, Hijano:2019qmi, Komatsu:2020sag, Chandorkar:2021viw}, it is expected to encode the S-matrix of the bulk theory. 
Away from the flat space limit, presently we are not equipped to answer this question because we do not have a notion of S-matrix that takes curvature corrections into account. In this paper, we seek to remedy this situation by defining an ``AdS S-matrix'' in ``momentum space'' using local bulk physics and describe how it computes the conformal correlators. To be precise we describe how it computes the Mellin amplitude in $1/\tR$ perturbation theory in the kinematical regime where the flat space limit is dominant. Here $\tR$ is the radius of AdS. In the large $\tR$ limit, our AdS S-matrix reduces to the flat space S-matrix. 

In order to make this discussion precise, we need to define the term ``flat-space limit'' carefully. We will do this when the bulk theory is weakly coupled. Throughout the paper we will restrict ourselves to this regime.
The flat space limit of the bulk theory is probed in a different way when at least one of the particles involved in the scattering process is massive compared to when all are massless. Let us first discuss the former case. In this case the dimension of the dual CFT operator(s) dual to the massive particle(s) needs to be scaled with $\tR$. This is because of the dictionary $m^2\tR^2=\Delta(\Delta-d)$. But because CFT doesn't know about $\tR$ \emph{per se}, the CFT observer can only  take large $\Delta$ limit without specifying whether it corresponds to large $\tR$ limit or large $m$ limit. 
Let us elaborate on what each of these limits actually means by considering an internal propagator $G_{BB}(X,Y)$ of a Witten diagram. For some external configurations, the large $\Delta$ limit of the $X,Y$ integral is dominated by the saddle point where $X=Y$ and for some, it is dominated by the saddle point where $X\neq Y$. The former can be thought of as the large $\tR$ limit because, the propagation of the virtual particle is over distances much smaller compared to $\tR$ i.e. in flat space. In the latter case, the internal particle has to propagate  along a geodesic over distance of $O(\tR)$, hence the large $\Delta$ limit effectively ends up being a large $m$ limit.  
Thus, for those external cross-ratio or momenta configurations for which \emph{every} propagator is dominated by the large $\tR$ limit as opposed to the large $m$ limit, the bulk process can be approximated to be happening in a small flat space  elevator. These are the configurations which produce the flat space S-matrix in the large $\Delta$ limit. This is what we call the flat space limit. For any other configuration, at least one bulk propagator is extended over $O(\tR)$ distance in the bulk and the large $\Delta$ limit does not correspond to flat space. 

When none of the particles, neither the external not the internal, involved in the scattering process is massless, the CFT observer can't take the large $\Delta$ limit. In this case, the flat-space physics is obtained  by taking the so called bulk-point limit $z=\bar z$. The correlator is singular in this limit on the appropriate sheet and  the bulk S-matrix is recovered as the coefficient of this singularity \cite{Heemskerk:2009pn,Maldacena:2015iua}. This singularity has been analyzed only for contact diagrams and to some extent for the exchange diagram but one expects that the singularity to be present to all orders in perturbation theory and that it encodes the flat space S-matrix. Because, the S-matrix does not have a mass-scale, it is a homogeneous function of Mandelstam variables i.e. it is has the form
\begin{equation}
    S(s,t)=s^af(s/t).
\end{equation}
The degree of homogeneity $a$ is is related to the strength of the bulk-point singularity and the function $f(x)$ is the coefficient of this singularity which is also a function of a single variable. At the bulk-point singularity, the boundary insertion points are null-separated from the bulk-point where the scattering takes place. To summarize
\begin{itemize}
    \item {\bf Large $\Delta$ limit}: When at least one massive particle is involved in the process, the flat space S-matrix is potentially obtained by taking large $\Delta$ limit.
    \begin{itemize}
        \item {\bf Flat space limit}: In some range of parameter space, the bulk process takes place in a small elevator in AdS and the large $\Delta$ limit corresponds to the flat space limit.
        \item {\bf Large $m$ limit}: In the complementary region of the parameter space, this is not true and there exist at least one internal geodesic propagation over distances of $O(\tR)$. 
    \end{itemize}
    \item {\bf Bulk-point limit}: When all of the particles in the scattering process are massless then the flat space S-matrix is expected to appear as the coefficient of the singularity at $z=\bar z$.
\end{itemize}   
The difference between the flat space limit for massless and massive particles is emphasized in the paper \cite{Hijano:2019qmi,Li:2021snj}.

In this paper, we consider the case of scattering of massive particles in the bulk i.e. we take large $\Delta$ limit of the CFT$_{d}$ correlators where $\Delta$ is the dimension of the operators involved in the scattering process. We will not identify the region in the parameter space where the large $\Delta$ limit corresponds to flat space limit but rather only assume that such a region exists and restrict ourselves to be in that region. We define a bulk observable that we call the AdS S-matrix. It has the following desirable properties: 
\begin{itemize}
    \item The AdS S-matrix can be computed using the local bulk-physics in $1/\tR$ perturbation theory, in particular using a set of Feynman-like rules directly in ``momentum'' space.
    \item It reduces to the flat space S-matrix in the large $\tR$ limit and the above-mentioned Feynman-like rules reduce to the usual Feynman rules.
    \item It can be used to compute Mellin amplitude in $1/\tR$ perturbation theory. As remarked earlier, this works in the regime of parameter space where the large $\Delta$ limit of the correlator corresponds to the flat space limit.
\end{itemize}
The word momentum in the first point and also in the first paragraph is in quotes because AdS does not have any translational symmetry. However, as we discuss below, we define momentum-like variable in a conformally covariant way which has the property that it reduces to the eigenvalue of the actual momentum generator in the flat space limit. 

The isometry group of anti-de Sitter space, say of AdS$_{d+1}$ is $SO(d,2)$. Let $M_{AB}, A,B=0,\ldots,d+1$ be the generators of of this group. We embed the AdS into the ${\mathbb R}^{d,2}$ space parametrized by coordinates $\eX^A$ as $\eX\cdot \eX=-\tR^2$. 
We divide the symmetry generators into two groups, $M_{ab}, a,b=0,\ldots,d$ that are the generators of the $SO(d,1)$ subgroup and the remaining $M_{a}=M_{a,d+1}$. Under the Inonu-Wigner contraction $M_a = \tR \, \pP_a$ with $\tR\to \infty$, the algebra $SO(d,2)$ reduces to the Poincare algebra. As we will argue later, this is the isometry algebra of the flat space near the point $(0,\ldots,0,\tR)\equiv \eC$ on the AdS hyperboloid. We  introduce new type of AdS variable $\eP$, similar to flat space momentum, that is Fourier conjugate to embedding space coordinate $\eX$. We refer to this conformally covariant variable $\eP$ as conformal momentum. We will show that when we take $\eP=(\pP_a,0)$, the bulk process localizes in the patch around the point $\eC$ and in this patch, $\pP$ precisely plays the role of the flat space momentum in the $\tR\to \infty$ limit.  We  refer to the variable $\pP$ as the AdS momentum.
The usefulness of this variable lies in efficiently computing the AdS S-matrix in $1/\tR$ perturbation theory and in relating it to more familiar CFT observable viz. the Mellin amplitude \cite{Mack:2009mi, Mack:2009gy, Penedones:2010ue, Fitzpatrick:2011ia} as we will show in the paper. 

Let us stress that neither the conformal momentum nor the AdS momentum is  same as the $d$-dimensional CFT momentum. 
The CFT correlation functions in momentum space have been widely studied \cite{Coriano:2013jba, Bzowski:2013sza, Bzowski:2019kwd, Coriano:2019sth, Isono:2018rrb, Isono:2019wex, Bautista:2019qxj, Gillioz:2019lgs, Albayrak:2020isk,  Gillioz:2020mdd, Jain:2022ujj}. These correlation functions, in appropriate limit,  have even been assigned an interpretation of a scattering amplitude in \cite{Gillioz:2019lgs}.  The momentum space CFT correlators have also been used to compute cosmological correlators or wave function coefficients in de Sitter space, see e.g. \cite{Arkani-Hamed:2018kmz, Baumann:2019oyu, Melville:2021lst}. It would be interesting to understand the connection between the these sets of variables. We will  not attempt do so in this paper.  

The AdS S-matrix is defined by using the Fourier transform of the position space correlator in the embedding space, supported only over the null cone. We will show that this $\eP$-space correlator $\hat G(\eP_i)$ evaluated on the slice $\eP_{i,{d+1}}=0$ reduces to the flat space S-matrix in the large $\tR$ limit. In particular it is proportional to momentum conserving delta function $\delta(\sum \pP_i), i=1,\ldots,4$ at leading order. As the translational symmetry is broken by the $1/\tR$ corrections, $\hat G(\eP_i)|_{\eP_{i,{d+1}}=0}$ will certainly have pieces that are proportional to the derivatives of $\delta(\sum \pP_i)$ at sub-leading orders in $1/\tR$. However, after fixing a certain ambiguity in the derivative expansion around $\delta(\sum \pP_i)$, we define the AdS S-matrix $S(\pP_i)$ to be the term that is proportional to $\delta(\sum \pP_i)$. It might seem that, in doing so, we lose information that is present the pieces proportional to the derivatives of  $\delta(\sum \pP_i)$ but that is not so. We will show how the conformal Ward identities can be used to reconstruct the full $\hat G(\eP_i)$ using only the AdS S-matrix in $1/\tR$ perturbation theory. This shows that our AdS S-matrix contains all the $1/\tR$ perturbative information in the conformal correlator.

In order to define the AdS S-matrix from CFT correlators, we need a type of LSZ prescription to construct on-shell states. As we will discuss in the paper,  the extrapolate dictionary of AdS/CFT provides us with one. In \cite{Komatsu:2020sag}, a way to compute flat space S-matrix from position space conformal correlator was proposed. This proposal was proved in \cite{Li:2021snj} by preparing fixed ``momentum'' states in the bulk using HKLL formula as in \cite{Hijano:2019qmi}.
Our prescription agrees with \cite{Komatsu:2020sag} at leading order. In addition to offering a symmetry based understanding of bulk momentum, the novelty of our prescription is that it allows for computation $1/\tR$ corrections which can be used to compute the full conformal correlator in perturbation theory and for relating the AdS S-matrix and the Mellin amplitude. 

The rest of the paper is organized as follows. In section \ref{fourier}, we introduce the covariant $\eP$-variables in AdS as Fourier conjugate to embedding space coordinates. We then extend the definition to boundary conformal field theory. After computing the CFT two point function $\eP$-space, we show that after specializing to $\eP=(\pP_a,0)$, the bulk contact interaction  localizes in a small patch near $\eC=(0,\ldots,0,\tR)$. In section \ref{s-matrix}
we first compute the boundary to bulk propagator and and show that $\pP_a$ indeed plays the role of the local momenta at $\eC$. We will define the AdS S-matrix using contact diagrams in the bulk and show how the $\eP$-space correlator $\hat G(\eP)$ is constructed from it in $1/\tR$ perturbation theory using conformal Ward identities. We develop Feynman-like rules to compute the AdS S-matrix in section \ref{feynman}. There we also compute the AdS S-matrix for scalar exchange and bubble diagrams explicitly to the first non-trivial sub-leading order in $1/\tR$. In section \ref{mellin}, we develope a dictionary between Mellin amplitude and the AdS S-matrix in $1/\tR$ perturbation theory. The paper is supplemented with two appendices which contain calculations pertinent to some of the results used in the bulk of the paper.

\subsection*{Notation}
We use uppercase calligraphic letters such as $\eX^A, \eZ^A, \eP^A$  to denote conformally covariant quantities i.e. $d+2$ dimensional quantities on which the conformal symmetry acts linearly. We use uppercase letters such as $\pX^a,\pZ^a$  to denote the $d+1$ dimensional vectors that live in the tangent space to the AdS at the bulk point $\eC=(0,\ldots, 0,\tR)$ and $\pP^a$ to denote the first $d+1$ components of $\eP$ after setting $\eP^{d+1}=0$.

\section{Fourier transform to momentum space}\label{fourier}
Conformal symmetry  is realized linearly in the so called embedding space. For a Lorentzian CFT in $\RR^{d-1,1}$, the embedding space is $\RR^{d,2}$. We will take the signature of the flat metric in the embedding space to be $g_{AB}={\rm diag}(-1,+1,\ldots, +1, -1)$, $A=0,\ldots, d+1$. Letting  $\eX^A$ to be the embedding space coordinates, the conformal generators are
\begin{equation}
    M_{AB}=\eX_A\frac{\partial}{\partial \eX^B}-\eX_B\frac{\partial}{\partial \eX^A}.
\end{equation}
The  $d$ dimensions are CFT are identified with the projective null cone i.e. $\eX_A \eX^A=0$ subjected to the identification $\eX^A\sim \lambda \,\eX^A$. 
The AdS space is the hyperboloid $\eX_A \eX^A=-\tR^2$. Here $\tR$ is known as the radius of AdS. 

Whenever the flat space limit is well-defined, we expect the bulk interactions to occur in a small patch of AdS. In the large $\Delta$ limit, if we further take $\eP_i$'s of the external operators to be of the form $\eP_i=(\pP_i,0)$ i.e. when we take $\eP_{i,d+1}=0$ then we will show that the patch in question is around the point $\eC=(0,\ldots,0,\tR)$. We denote the coordinates around $\eC$ by $\pX_a$. Concretely, we will take $\eX=(\pX_a,(\tR^2+\pX\cdot \pX)^{1/2})$. When the patch is approximated by flat space, the conformal generator $M_{ab}$  becomes  Lorentz transformation and the remaining, $M_{a,d+1}\equiv M_a$ becomes $\tR$ times the momentum. As remarked earlier, the variable $\pP$ will precisely turn out to be this local momentum. Reader might be worried that we are losing generality by taking $\eP_i$'s to be of this special form but that is not so. This merely fixes a conformal frame for the $\eP$-space correlator. As we will show later, the entire kinematic regime can be explored even with this choice. For any other choice of frame also the bulk process would be localized at some point but not necessarily at $\eC$. We prefer to localize the bulk process at $\eC$ as  it is technically convenient. Also it makes a direct connection with the dictionary between CFT coordinates and bulk momenta proposed in \cite{Komatsu:2020sag}. 

\subsection{Momentum space for AdS}\label{AdS-momentum}
In this section, we introduce the conformal momentum $\eP$ and the AdS momentum $\pP$. The $\pP$-variables are desired to have the property that they reduce to eigenvalues of $M_a$ in the flat space limit. Note that this property can only be expected of the flat space limit because away from it, $M_a$'s do not commute with each other and hence can not be simultaneously diagonalized. Indeed their commutation relation is
\begin{equation}
    [M_a,M_b]=M_{ab}.
\end{equation}
We demand that the conformal symmetry acts linearly on these variables. This means that naturally they should be Fourier dual to embedding space coordinates. Let $H(\eX_i)$ be a position space correlation function in the bulk where $\eX_i$ labels insertion points of $i$th operator $\Phi_i$. We define the $P$-variables with the Fourier transform in embedding space restricted to the AdS hyperboloid,
\begin{equation}\label{ads-fourier}
    {\tilde H}(\eP_i)= \int \prod_{i} d\eX_i\, \delta(\eX_i^2+\tR^2)\,e^{i\sum_i \eP_i\cdot \eX_i} H(\eX_i).
\end{equation}
Due to the manifest covariance of the above expression, the conformal generators do act linearly on $\eP_A$ variables as desired,
\begin{equation}
    M_{AB}=\eP_A\frac{\partial}{\partial \eP^B}-\eP_B\frac{\partial}{\partial \eP^A}.
\end{equation}
Because the integration is restricted to the hyperboloid, the Fourier transform has the property,
\begin{equation}\label{hyperboloid-constraint}
    \Big(\frac{\partial}{\partial \eP_i} \cdot \frac{\partial}{\partial \eP_i}- \tR^2 \Big) \tilde H(\eP_i)=0
\end{equation}
for all $i$. Instead of thinking of $\tilde H(\eP_i)$ as a function of $d+2$ variables $\eP_{i,A}$ constrained as above, it is more convenient to look at the slice $\eP_{i,d+1}=0$. Then $\tilde H(\eP_i)$ is an unconstrained function of $d+1$ variables $\eP_{i,a}$, $a=0,\ldots, d$. This vector plays the role of the unconstrained bulk ``momentum''. Just like in the position space, we will denote the $d+1$ dimensional $P$-vector as $\pP_a$. The action of 
conformal generators  on $\pP_a$ can be readily worked out,
\begin{equation}
    \label{ads-rep}
    M_{ab}=\pP_a\frac{\partial}{\partial \pP^b}-\pP_b\frac{\partial}{\partial \pP^a},\qquad \quad M_{a,d+1}=\pP_a\frac{\partial}{\partial \eP^{d+1}}-\eP_{d+1}\frac{\partial}{\partial \pP^a}=i\,\pP_a\,\sqrt{\tR^2-\frac{\partial}{\partial \pP} \cdot \frac{\partial}{\partial \pP}}.
\end{equation}
For the last equality we have used $\eP_{d+1}=0$ and the equation \eqref{hyperboloid-constraint}. It is easy to check that these generators indeed satisfy the conformal algebra. It is instructive to look at the action of these generators at the point $\eC$ where the bulk process is localized. The generators $M_{ab}$ become Lorentz transformations while 
\begin{equation}
    M_{a,d+1}\to \pP_a \eC_{d+1}=i\,\tR\, \pP_a.
\end{equation}
This is the Inonu-Wigner contraction of the conformal algebra to the Poincare algebra in the flat space patch near $\eC$. A similar contraction in position space gives, $M_{a,d+1}\rightarrow-\tR\partial/\partial \pX^a$. As promised  $\pP_a$ is indeed momentum in the patch near $\eC$ in the large $\tR$ limit.

In this patch, the Fourier transform \eqref{ads-fourier} can be written in terms of the local coordinates as 
\begin{equation}
    {\tilde H}(\pP_i)= \int \prod_{i} \frac{ d^{d+1}\pX_i}{\sqrt{\tR^2+\pX_i^2}} \,e^{i\sum_i \pP_i\cdot \pX_i} \,  H(\pX_i,\sqrt{\tR^2+\pX_i^2}).
\end{equation}
Here, we have done the $\eX_{d+1}$ integral using the delta function in equation \eqref{ads-fourier} and picked a particular branch for the solution that is in the neighborhood of the point $\eC$ (and not $-\eC$) and have also set $\eP_{d,+1}=0$ in equation \eqref{ads-fourier} and obtained the bulk correlation function directly in $\pP_i$s. 

\subsection{Momentum space for CFT}\label{CFT-momentum}

The function $\tilde H(\pP_i)$ is defined for general off-shell momenta. This is expected since it is simply the Fourier transform of bulk correlation function and not quite the ``S-matrix''. In order to compute the ``S-matrix" in the bulk, we have to perform the appropriate LSZ reduction. In flat space, the notion of S-matrix is \emph{a priori} well-defined and the standard LSZ prescription takes us from the position space correlation function to this \emph{a priori} well-defined observable. In AdS, on the other hand, the meaning of an on-shell S-matrix is less clear because asymptotic states are not as straightforwardly defined. Fortunately there \emph{is} a useful observable which is intimately related to what one would like to call the AdS S-matrix - the CFT correlation function. In this section, we define a Fourier transform that takes the CFT position space correlation function to momentum space and see that this  Fourier transform is indeed defined only for momenta that are ``on-shell''.  It is this momentum space correlation function that we would like to call the AdS S-matrix.\footnote{We will actually define the AdS S-matrix  to be the part of the momentum space correlator that is proportional to the momentum conserving delta function. For details see section \ref{S-matrix-contact}.} In the flat space limit, the AdS S-matrix goes over to the ordinary S-matrix in Minkowski space. In this way, the CFT operator insertions prepare on-shell asymptotic states for scattering in AdS.  We will describe this now.

We would like to define Fourier transform to $\eP$-space of CFT correlation functions that is compatible with the Fourier transform \eqref{ads-fourier} in AdS. The CFT correlation function in position space can be computed from the AdS correlation function using the so called ``extrapolate dictionary'' i.e.
\begin{equation}\label{extrapolate}
    \CO(\eW)= \lim_{\alpha \to \infty} \alpha^{\Delta}\Phi(\eX=\alpha \eW+\ldots).
\end{equation} 
Here, the embedding space vector $\eW$ obeys $\eW^2=0$ labelling a point on the boundary and $\eX$ is a point in AdS obeying $\eX^2=-\tR^2$. The $\ldots$ are the $O(\alpha^0)$ terms required for the condition $\eX^2=-\tR^2$ to hold. For example, $\ldots$ could be some vector $\eY$ that obeys $\eY^2=-\tR^2$ and $\eY\cdot \eW=0$. This lets us naturally restrict the bulk Fourier transform to the boundary by taking the limit of the AdS position space correlators to the boundary as in equation \eqref{extrapolate}. Explicitly,
\begin{align}\label{cft-fourier}
   % {\tilde G}(\eP_i)&\equiv \lim_{\alpha\to \infty} \alpha^{\sum_i \Delta_i}\int \prod_{i} d\eX_i \,\delta(\eX_i^2+\tR^2)\, e^{i\frac{1}{\alpha}\sum_i \eP_i\cdot \eX_i} G(\eX_i)  \,\,\qquad \eX_i=\alpha \eW_i+\ldots \notag\\
   % \Rightarrow \quad 
    {\tilde G}(\eP_i)&= \int \prod_{i} d\eW_i \delta(\eW_i^2) \,e^{i\sum_i \eP_i\cdot \eW_i} G(\eW_i) \,\,.
\end{align}
This definition looks almost the same as the one in equation \eqref{ads-fourier} except that the Fourier transform is on the null-cone rather than on the hyperboloid. As a result, the Fourier transform is a harmonic function of $\eP_i$ i.e. 
\begin{equation}\label{null-constraint}
    \frac{\partial}{\partial \eP_i} \cdot \frac{\partial}{\partial \eP_i} \tilde G(\eP_i)=0.
\end{equation}
This equation is the boundary analogue of the AdS constraint \eqref{hyperboloid-constraint}. As before, this allows us to set $\eP_{d+1}=0$ and write $\eP=(\pP_a,0)$ just like in the case of AdS.

Thanks to the homogeneity of the CFT correlation function, the Fourier transform ${\hat G}(\eP_i)$ is also homogenous. 
\begin{align}\label{homog}
    &G(\lambda_i \eW_i)=\Big(\prod_i \lambda_i^{-\Delta_i}\Big) G(\eW_i) \quad \Rightarrow \quad \tilde G(\lambda_i \eP_i)=\Big(\prod_i \lambda_i^{\Delta_i-d}\Big) \tilde G(\eP_i)\notag\\
     &\hat G(\eP_i)\equiv  \,\,\tilde G(\eP_i)\,\,\Big(\prod_i |\eP_i|^{d-\Delta_i}\Big).
\end{align}
Here, the new function $\hat G(\eP_i)$ is only a function of the unit vector $\eP_i/|\eP_i|$. Since it does not depend on $|\eP_i|$, we can gauge fix $|\eP_i|$. For $\Delta\neq d$ i.e. for massive particles in the bulk, we choose $\eP_i\cdot \eP_i=-m_i^2=\pP_i\cdot \pP_i$. We will justify this choice using conformal Casimir near equation \eqref{casimir1} and also from the boundary to bulk propagator in section \ref{lsz}.  With this choice, the Fourier transform of CFT correlation functions defined in equation \eqref{cft-fourier} has effectively become a function of on-shell AdS momenta $\pP_i$. Thus the on-shell asymptotic states are constructed by supporting the Fourier transform only over the boundary.

Also note that, thanks to homogeneity \eqref{homog}, the Fourier transform of CFT correlation functions can also be written as
\begin{equation}\label{cft-fourier-2}
   \tilde G(\eP_i)=\int \prod_{i} d\eW_i\,\delta(\eW_i^2) \,\Big(\prod_i \Gamma(d-\Delta_i)(i\eP_i\cdot \eW_i)^{\Delta_i-d}\Big)  G(\eW_i).
\end{equation}
Here we have used the Cahen-Mellin formula $e^{-x}=\frac{1}{2\pi i}\int ds\,\Gamma(s)\, x^{-s}$ where the contour is parallel to the imaginary axis and the fact that the boundary $d\eW_i$ integral is non-zero only when it is conformal i.e. the integrand has degree $-d$ in $\eW_i$. The advantage of this expression over \eqref{cft-fourier} is that the integral is homogeneous in $\eW_i$. Usual technique of evaluating conformal integrals by gauge fixing are readily applicable to \eqref{cft-fourier-2}. 

\subsection{Two point function}
Let us familiarize ourselves with this Fourier transform by computing the CFT two point function in momentum space. There are multiple ways to compute it. We can either compute it directly by Fourier transforming position space two point function or by using conformal Ward identities. We present the latter method here and the former in appendix \ref{exact-prop}.
 
As is well known, the CFT two  point function is completely fixed by conformal symmetry. For simplicity, consider the two point function of two scalar operators with conformal dimension $\Delta_1$ and $\Delta_2$. The two point function is non-zero only when $\Delta_1=\Delta_2$ and is given by, 
\begin{equation}
    \langle \CO(\eW_1) \, \CO(\eW_2)\rangle =G(\eW_1,\eW_2)= (-\eW_1\cdot \eW_2)^{-\Delta_1}.
\end{equation} 
This is a unique function of $\eW_1$ and $\eW_2$ that is conformally invariant and has the correct homogeneity in both the variables. The two-point function in $P$-space must also have the homogeneity given in equation \eqref{homog}. This fixes the form of the momentum space function,
\begin{equation}
    \hat G(\eP_1,\eP_2)=  f(\chi)\qquad \chi\equiv \frac{(\eP_1\cdot \eP_2)^2}{\eP_1^2 \eP_2^2}.
\end{equation}
At first glance, it seems that, unlike in position space, the correlator in momentum space is not completely  fixed by conformal symmetry and contains an undetermined function of the ``cross-ratio" $\chi$. However, this is not the case. The function $f(\chi)$ is seemingly undetermined because we have not taken all the constraints into account, in particular we have not taken into account the null-space condition \eqref{null-constraint}. The correlator obeys equation \eqref{null-constraint} for both the momenta $\eP_1$ and $\eP_2$ but as it is symmetric between them, it suffices to impose only one condition, say corresponding to $\eP_1$. This gives a second order differential equation,
\begin{equation}\label{hypergeom}
\chi(1-\chi) \ f''(\chi) + \frac{1}{2} (1-(d+2)\chi) \ f'(\chi) + \frac{1}{4} \Delta(\Delta-d) \ f(\chi) =0
\end{equation}
The solution is a linear combination of Gauss hypergeometric functions,
\begin{equation}\label{exact-2}
f(\chi) = \alpha \ _{2}F_{1}\Big(\frac{d-\Delta}{2}, \frac{\Delta}{2}, \frac{1}{2} , \chi \Big) + \beta \ \sqrt{\chi} \ _{2}F_{1}\Big(\frac{d-\Delta+1}{2}, \frac{\Delta+1}{2}, \frac{3}{2} , \chi \Big).
\end{equation}
The first solution in even in $\sqrt{\chi}$ and hence under the operation $\eP_1\to e^{i\pi }\, \eP_1, \eP_2\to \eP_2$ and the second is odd. 

That there are two independent solutions for $\hat G(\eP_1,\eP_2)$ may seem in contradiction with the fact that the $G(\eW_1,\eW_2)$ is uniquely determined by conformal symmetry, but it is not. The two solutions are  due to the ambiguity in defining the Fourier transform \eqref{cft-fourier}, in particular its support of integration. Let us first understand this in Euclidean space. The fourier transform \eqref{cft-fourier} is supported over the entire null locus $\eW^2=0$ but we could have restricted it to either the future or the past null cones as conformal symmetry acts on them separately. The solution \eqref{exact-2} is computed only using conformal symmetry and without using a definite choice as made in \eqref{cft-fourier}. This is why we find two linearly independent solutions corresponding to restricting the integral to either future or past null-cones. For the concrete choice of integration domain to be the entire null locus i.e. the union of future and the past null-cones, it is easy to see that $\hat G(\eP_1,\eP_2)$ is even under $\eP_1\to e^{i\pi}\,\eP_1, \eP_2\to \eP_2$ and hence corresponds to setting $\beta=0$ in \eqref{exact-2}. 
\begin{align}
    \hat G(e^{i\pi}\eP_1,\eP_2)&=|e^{i\pi} \eP_1|^{d-\Delta}|\eP_2|^{d-\Delta} \int  d\eW_1 d\eW_2 \,e^{i(e^{i\pi}\eP_1\cdot \eW_1+\eP_2\cdot \eW_2)} G(\eW_1,\eW_2) \,\,\delta(\eW_1^2)\delta(\eW_2^2)\notag\\
    &=|e^{i\pi}\eP_1|^{d-\Delta}|\eP_2|^{d-\Delta}(-1)^d \int  d\eW'_1 d\eW_2 \,e^{i(\eP_1\cdot \eW'_1+\eP_2\cdot \eW_2)} G(e^{i\pi}\eW'_1,\eW_2) \,\,\delta(\eW_1^{'2})\delta(\eW_2^2)\notag\\
    &=\hat G(\eP_1,\eP_2).
\end{align}
Here we have made the substitution $e^{i\pi}\eW_1=\eW'_1$ in the second line and have used the symmetry of the integration domain under $\eW_1\to -\eW_1$. In appendix \ref{exact-prop}, we have computed the values of $\alpha$ and $\beta$ when the range of integration is taken to be only the positive null cone for both variables $\eW_i$'s.

In Lorentzian space also, the minus sign of $\eP_1$ can be absorbed into $\eW_1$. As a result $\eW_1\cdot\eW_2\to -\eW_1.\eW_2$. In other words, changing the sign of $\eP_1$ changes the space-like separation between operators to time-like and vice versa. If we define the Fourier transform so that it has support everywhere on the Lorentzian cylinder, the same argument as above shows us that $\hat G(\eP_1,\eP_2)$ is an even function under $\eP_1\to e^{i\pi}\,\eP_1, \eP_2\to \eP_2$ just like in the Euclidean space. 

\subsubsection{In the large $\tR$ limit}\label{two-pt}
Evaluating the two point function in the large $\tR$ limit gives more insight into the relation of conformal momentum space observables to those in flat space. This is what we will do now.
In this section we compute the Fourier transform \eqref{cft-fourier} of the CFT two point function of two scalar operators.
\begin{equation}
    {\tilde G}(\eP_1,\eP_2)= \tR^{2(\Delta-d)}\int d\eW_1 \int   d\eW_2 \,e^{i \,\tR \,(\eP_1\cdot \eW_1+\eP_2\cdot \eW_2)} (\eW_1\cdot \eW_2)^{-\Delta}\, \delta(\eW_1^2)\,\delta(\eW_2^2).
\end{equation}
Here we have scaled $\pW_i\to \tR \pW_i$ so that $\tR$ appears in the exponent.
In order to compute the integral we will first solve the null constraint $\eW_i^2=0$ as $\eW_i=(\pW_i,|\pW_i|)$ where $\pW_i$ is a $d+1$ vector. The integral becomes,
\begin{equation}\label{CFT-2pt}
    {\tilde G}(\pP_1,\pP_2)= \tR^{2(\Delta-d)}\int \frac{d\pW_1}{|\pW_1|} \int  \frac{d\pW_1}{|\pW_1|} \,e^{i \,\tR \,(\pP_1\cdot \pW_1+\pP_2\cdot \pW_2)} (\pW_1\cdot \pW_2-|\pW_1||\pW_2|)^{-\Delta}.
\end{equation}
Here we have set $\eP_{d+1}=0$ as discussed above.
The saddle-point equations are
\begin{equation}
    i\pP_1=m\frac{\pW_2-\frac{|\pW_2|}{|\pW_1|} \pW_1}{\pW_1\cdot \pW_2-|\pW_1||\pW_2|}, \qquad i\pP_2=m\frac{\pW_1-\frac{|\pW_1|}{|\pW_2|} \pW_2}{\pW_1\cdot \pW_2-|\pW_1||\pW_2|}. 
\end{equation}
These equations can be solved for $\pW_i$ as 
\begin{equation}\label{saddle1}
    \pW_1^*= -\pW_2^*=-i\frac{m}{\pP_1^2} \pP_1.
\end{equation}

At the saddle point, we compute the value of the  generator $M_a$ on the boundary.
\begin{equation}\label{casimir1}
    M_a=i\tR \eW_{d+1}^* \, \pP_a= i\tR |\pW_a^*| \pP_a = -\frac{m\tR}{|\pP|} \pP_a.
\end{equation}
From this equation, it is clear that we must fix the gauge $\pP^2=-\,m^2$ in order to interpret the variables $\pP_a$ as bulk momenta $-iM_a/R$. Happily, this choice of $|\pP_a|$ also corresponds to on-shell states as remarked near equation \eqref{homog}. This in particular means that the integral has support only for $\pP_1=-\pP_2$ i.e. it is proportional to $\delta(\pP_1+\pP_2)$ as expected of a two point function in momentum space.\footnote{The reason we got a unique answer is because we tacitly chose the support of integration of $\eW_1$ and $\eW_2$ to be the future null-cone. This is done while solving the $\eW_i^2=0$ constraint as $(\pW_i,+|\pW_i|)$. If we had chosen $\eW^{d+1}=-|\pW|$ for one of the points then we would have obtained a solution that is proportional to $\delta(\pP_1-\pP_2)$.} 

This choice of the gauge can also be argued more generally by matching the flat space limit of the conformal Casimir with the Poincare Casimir in the flat space. In the large $\tR$ limit, the generators $M_a$ should be proportional to $\pP_a$, say $M_a=c \tR\, \pP_a$ as that is the only other $d+1$ vector that can appear on the right hand side. Then the constant of proportionality is determined by equating the Casimirs
\begin{equation}
    M_a^2+M_{ab}^2=\Delta(\Delta-d)+\ell(\ell+d-2) \qquad \Rightarrow \qquad c^2 \pP_a^2=m^2.
\end{equation}
The solution $c^2=m^2/\pP^2$ agrees with equation \eqref{casimir1}. 
In this derivation, we have used the fact that the contribution of the Lorentz generators $M_{ab}$ to the Casimir is the spin part $\ell(\ell+d-2)$.

Another lesson from this exercise, is that, in the flat space limit, the two point function in momentum space is obtained from the position space by effectively substituting 
\begin{equation}\label{prescription}
    \pW_a=i \pP_a/m.
\end{equation}
This is the momentum space prescription that in conjectured in \cite{Komatsu:2020sag}\footnote{In \cite{Komatsu:2020sag}, the authors work with Euclidean AdS and fix the gauge $|\pW|=1$. Due to this, their prescription is $\pW_0=-\pP_0/m,\, \pW_i=i\pP_i/m$.}. Here we see that this mysterious prescription of evaluating the position space correlation function at an imaginary position space argument is simply a consequence of Fourier transform getting dominated by a complex saddle. Moreover, our discussion makes it clear that $\pP_a$ are precisely bulk-momenta in the flat space limit.

In \cite{Komatsu:2020sag}, the authors demonstrate the relation between the momentum space correlator defined using equation \eqref{prescription} to the Mellin amplitude in the strict large $\tR$ limit. 
The advantage of defining the momentum space using the Fourier transform \eqref{cft-fourier} as opposed to using the prescription in \cite{Komatsu:2020sag} is that the relation between momentum space observable and the Mellin amplitude can be established beyond the flat space limit. We will compute this dictionary in section \ref{mellin}. 
We will also develop a simple set of Feynman-like rules in momentum space to compute Witten diagrams in $1/\tR$ perturbation theory and using the aforementioned dictionary we can then compute the Mellin amplitude in $1/\tR$ perturbation theory.

\subsection{Localization at the bulk point}\label{bulk-loc}
As we have discussed in the introduction, for a range of cross-ratios, the bulk process is localized to a point in AdS. In this section we will show that after specializing to $\eP=(\pP_a,0)$ the bulk point in question is $\eC=(0,\ldots,0,\tR)$. We will show this for the simplest contact diagram, namely $\phi^4$. The argument is shown to be valid for general correlation functions in section \ref{mellin} using Mellin space representation.

\begin{equation}\label{contact-position}
    G_{\phi^4}(\eW_i)=\int d\eZ\,\delta(\eZ^2+\tR^2) \prod_i(\eW_i\cdot \eZ)^{-\Delta}.
\end{equation}
This integral has been analyzed in \cite{Komatsu:2020sag} for $\eW_i=(\pW_i,1)$ with $\pW_i=i\pP_i/m, \pP_i\in {\mathbb R}$. In general it is difficult to compute the saddle point of the $\eZ$ integration  $\eZ_*$ as a function of $\pP_i$. However, if we assume that $\pP^0_i$ component for two of the four $\pP_i$ is positive and for the other two it is negative then the authors show that the integral is exponentially sub-leading away from the locus $\sum_i\pP_i=0$ i.e. it is proportional to $\delta(\sum_i \pP_i)$ and on this locus, the saddle point of $\eZ$ integration is at $\eC$.\footnote{The saddle point also exists at $-\eC$, but we focus on the point $\eC$ because we are interested in the scattering process taking place there.}

The $\eP$ space correlation function corresponding to the $\phi^4$ contact diagram is 
\begin{equation}\label{Pspace-contact}
    {\tilde G}_{\phi^4}(\eP_i)|_{\eP_i^2=-m_i^2}= \int \prod_{i} d\eW_i\, \prod_{i}\delta(\eW_i^2) \,e^{i\sum_i \eP_i\cdot \eW_i} \int d\eZ\,\delta(\eZ^2+\tR^2) \prod_i(\eW_i\cdot \eZ)^{-\Delta}.
\end{equation}
The $\eZ$ integral for $\eP$-space correlation function, after setting $\eP_i^{d+1}=0$,  also can be shown to localize precisely at the same point $\eC$ for the values of $\pP_i$'s that maximizes $\tilde G(\pP_i)$.\footnote{Setting $\eP_i^{d+1}=0$ is not essential. We can set them to be positive constants $c_i$ and the saddle point for $\eZ$ is unaffected.} This is exhibited in detail in appendix \ref{bulk-loc}. Here we describe a shortcut to understand the same using the analysis of \cite{Komatsu:2020sag} in position space.

Even though $\pP_i$ are not integration variables, to find the locus of $\pP_i$ that yields dominant integral we need to write variational equations for $\pP_i$ as if they were integration variables. Moreover, because  $\eP_i^{d+1}=0$, $\pP_i$ are constrained to obey $\pP_i^2=-m_i^2$. We impose this by including the factor 
\begin{equation}
    \int d\lambda_i \, e^{\lambda_i(\pP_i^2+m_i^2)}
\end{equation}
in the above expression. Varying with respect to $\pP_i$ yields,
\begin{equation}
    2\lambda_i \pP_i=i \pW_i.
\end{equation}
Recalling $|\pW_i|=1$, we get $\pW_i=i\pP_i/m_i$. This is precisely the prescription of \cite{Komatsu:2020sag}. After substituting $\pW_i=i\pP_i/m_i$ we get the integral \eqref{contact-position}. At this stage, the analysis of \cite{Komatsu:2020sag} applies and the $\eZ$ integral gets localized to $\eC$.

\section{External leg and the S-matrix}\label{s-matrix}
In this section, we compute the boundary to bulk propagator in momentum space and interpret it as the momentum space ``external leg'' factor for Witten diagrams. As showed in section \ref{bulk-loc}, the bulk process is localized in a small patch near $\eC$. That is why, in this section, we will mainly focus on computing the boundary to bulk propagator in $1/\tR$ perturbation theory with the bulk point being in the patch near $\eC$.

We will then define the ``AdS S-matrix'' with the help of a simple Witten diagram computation and discuss constraints of conformal Ward identities on momentum space correlation function. We then compute bulk to bulk propagator in momentum space.   

\subsection{LSZ prescription}\label{lsz}
Before we move to computing the boundary to bulk propagator in $1/\tR$ perturbation theory, let us evaluate it exactly. The position space boundary to bulk propagator for a field of dimension $\Delta$ is $(\eW\cdot \eX)^{-\Delta}$. The Fourier transform is
\begin{align}\label{boundary-bulk}
    {\tilde G}_{\partial B}(\eP;\eX)&= \int d\eW \,\,\delta(\eW^2) \,\,e^{i\sum_i \eP\cdot \eW}\, (\eW\cdot \eX)^{-\Delta}\notag\\
    &=|\eP|^{\Delta-d}\hat G_{\partial B}(\eta),\qquad \eta\equiv \frac{\eP\cdot\eX}{-i|\eP|}.
\end{align}
The function $\hat G_{\partial B}(\eta)$ can be fixed using the harmonicity condition \eqref{null-constraint} just like for the  boundary two point function. It then obeys the second order differential equation \ref{hypergeom} and we get,
\begin{equation}
    \tilde G_{\partial B}(\eta) = \alpha \ _{2}F_{1}\Big(\frac{d-\Delta}{2}, \frac{\Delta}{2}, \frac{1}{2} , \frac{\eta^2}{\tR^2} \Big) + \beta \ \eta \ _{2}F_{1}\Big(\frac{d-\Delta+1}{2}, \frac{\Delta+1}{2}, \frac{3}{2} , \frac{\eta^2}{\tR^2} \Big).
\end{equation}
The fact that there are two solutions is again related to the ambiguity in defining the support of the $\eW$ integral. If we integrate over both the null cones, we get a function that is even in $\eta$ i.e. with $\beta=0$.

\subsubsection{Large $\tR$ limit: Saddle point}
As remarked earlier, for us it is more useful to compute the boundary to bulk propagator in the flat space limit with the bulk point $\eX$ taken to be very close to $\eC$ i.e. $\eX=(\pX,\sqrt{\tR^2+\pX^2})$ with $\pX$ is $\CO(1)$.
\begin{align}
    {\hat G}_{\partial B}(\eP;\eX)&= \tR^{\Delta-d}\int d\eW \,\,\delta(\eW^2) \,\,e^{i\tR\sum_i \eP\cdot \eW}\, (\eW\cdot \eX)^{-m \tR}\notag\\
    &=\tR^{\Delta-d+1}\int \frac{d \lambda}{2\pi} \int d\pW \,e^{i\lambda\tR (\pW^2-1)}\,e^{i\tR\sum_i \pP\cdot \pW}\, (\pW\cdot \pX-\sqrt{\tR^2+\pX^2})^{-m \tR}
\end{align}
In the large $\tR$ limit, the last factor is approximated as $(-\tR)^{-m\tR}e^{m \pW\cdot \pX}$. It does not contribute to the saddle point as it does not scale exponentially in $\tR$.  The saddle point for the $\pW$ integration can be easily computed to be
\begin{equation}\label{bulk-bb-saddle}
    2i\lambda \pW_*=i\pP\qquad\Rightarrow\qquad \pW_*=i\pP/m.
\end{equation}
We have assumed $\pP^2=-m^2$ to get to the second equation. 
Equation \eqref{bulk-bb-saddle} is same as the relation between position and momentum that we have previously encountered in section \ref{bulk-loc}. Substituting this back into the integrand,
\begin{equation}
    \hat G_{\partial B}(\eP,\eX)= \alpha \, e^{i\pP\cdot \pX}
\end{equation}
where $\alpha$ is a constant. As expected, the boundary Fourier transform defined in \eqref{cft-fourier} has prepared plane-wave with momentum $\pP$ in the flat space patch near $\eC$ if we take $\pP$ to be on-shell. This is yet another way of justifying the LSZ prescription advocated below equation \eqref{homog}. We are interest in computing $1/\tR$ corrections to this leading order answer however it is cumbersome to do so for the saddle point integral. We will employ a different method to compute them more efficiently.

\subsubsection{Large $\tR$ limit: Differential equation}
Instead of the saddle point, the propagator can be computed in $1/\tR$ perturbation theory using the differential equation obeyed by the Green's function. We will set $\eP_{d+1}=0$ at the outset and with abuse of notation, write  $\tilde G_{\partial B}(\eP;\eX)=\tilde G_{\partial B}(\pP;\pX)$. It obeys the differential equation,
\begin{align}\label{EOM2pt}
    \big[-\nabla_{\pX}\cdot \nabla_{\pX} + m^{2} \big]  \tilde G_{\partial B}(\pP;\pX) &= 0 \notag\\
    \big[ -\partial_{\pX} \cdot \partial_{\pX} + m^2 - \frac{1}{\tR^2} ( \pX^{a} \pX \cdot \partial_{\pX} \partial_{\pX,a} + (d+1) \pX \cdot \partial_{\pX} ) \big]  \tilde  G_{\partial B}(\pP;\pX) &= 0 \notag\\
    \hat G''_{\partial B}(\eta) - m^2 \hat G_{\partial B}(\eta) +\frac{1}{\tR^2} ( \eta^2 \hat G''_{\partial B}(\eta) +\eta(d+1) \hat G'_{\partial B}(\eta)) &=0
\end{align}
where $\eta=i\pP\cdot\pX/|\pP|$ and $\hat G_{\partial B}(\eta)$ are as defined in equation \eqref{boundary-bulk}. From the last equation, it is clear that this equation has two solutions that go as $e^{\pm im\eta}=e^{\pm i\pP\cdot\pX}$ at leading order corresponding to positive and negative energy respectively. We pick the positive energy solution. The full form is written as
\begin{equation}\label{leg}
    \hat G_{\partial B}(\pP;\pX) \equiv g_{\partial B}(\eta)e^{im\eta}= \Big(1+\frac{1}{\tR^2} g_{\partial B}^{(1)}(\eta)+\ldots\Big)e^{im\eta}.
\end{equation}
First correction in this series are
\begin{align}
    g_{\partial B}^{(1)}(\eta)&=\frac{id}{4m}\eta+\frac{d}{4}\eta^2+\frac{i m}{6}\eta^3.
\end{align}
Equation \eqref{leg} is the external-leg factors for Witten diagrams in momentum space. At leading order, it is same as the leg factor for Feynman diagrams and $1/\tR$ corrections are straightforwardly computed.

\subsubsection{Action of $M_a$}
In order to impose conformal Ward identities on the momentum space correlation function, it is useful to evaluate the action of conformal generators on the external leg factors. These factors are manifestly invariant under $M_{ab}$ but as we have treated $d+1$ direction differently, the invariance under $M_a$ is non-trivial. Let us compute the action of $M_a$ on the external label $\pP$.

The $M^{a}$ operator in $\eP$ space is,
\begin{equation}
M^{a} =\pP_a\frac{\partial}{\partial \eP^{d+1}}-\eP_{d+1}\frac{\partial}{\partial \pP^a}=\pP_a\frac{\partial}{\partial \eP^{d+1}}.
\end{equation}
In the last equation, we have set $\eP_{d+1}=0$. The action of $M_a$ of $\tilde  G(\eP;\eX)$ can be easily evaluated.
\begin{equation}\label{Ma-action}
    \frac{1}{i\tR}\,M_a\,\tilde  G(\eP;\eX)|_{\eP_{d+1}=0}=\pP_a\Big(1+\frac{1}{\tR^2}\frac{1}{4m^2}\Big(d-2d\,\,\pP\cdot\partial_{\pP}-2\pP_a\pP_b\partial_{\pP}^a\partial_{\pP}^b-2m^2\partial_{\pP}\cdot\partial_{\pP}\Big)+\ldots\Big)\tilde G(\pP;\pX).
\end{equation}
Higher order corrections to $M_a$ can be computed straightforwardly. 
It is also easy to check that $M_a$ commutes with $\pP\cdot \pP$ in $1/\tR^2$ perturbation theory. In fact the conformal Casimir $M_{AB}M^{AB}$ turns out to be $\tR^2 \pP^2=-m^2\tR^2$ to all orders in perturbation theory as expected.  

\subsection{S-matrix with contact diagrams}\label{S-matrix-contact}
\begin{figure}[h]
    \begin{center}
    \includegraphics[scale=1.0]{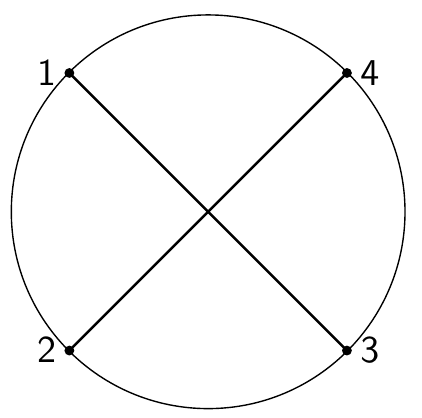}
    \end{center}
    \caption{Contact Witten diagram for $\phi^4$ interaction.}
    \label{contact-witten-dia}
\end{figure}

Equipped with the momentum space prescription for external leg, we are now ready to compute arguably the simplest type of bulk S-matrices: resulting from the tree level contact interactions in AdS. To begin with we consider the interaction $\phi^4$ of a scalar field with mass $m$. The bulk S-matrix is given by the Witten diagram given in figure \ref{contact-witten-dia}.
\begin{equation}\label{contact1}
    \hat G(\pP_i)=\int \frac{\tR\, d\pX}{\sqrt{\tR^2-\pX^2}}  \prod_{i=1}^4 g_{\partial B}\Big(\frac{\pP_i\cdot \pX}{m_i}\Big)e^{i \pP_i\cdot \pX}
\end{equation}
Due to presence of $\pX$ in the integrand, the S-matrix is not proportional to the momentum conserving delta function. This is expected as the translational symmetry of the flat space limit of AdS is broken by $1/\tR^2$ corrections. 
Equation \eqref{contact1} can be expanded according to $\pX$ homogeneity to have the form
\begin{align}\label{general-form}
    \hat G(\pP_i)\prod_i
    &=\int d\pX \Big(S^{(0)}(\pP_i)+[S^{(1)}(\pP_i)]^{a_1} i\pX_{a_1} +[S^{(2)}(\pP_i)]^{a_1 a_2} i\pX_{a_1} i\pX_{a_2}+\ldots \Big)e^{i(\sum \pP_i)\cdot \pX}\notag\\
    &=\Big(S^{(0)}(\pP_i)+[S^{(1)}(\pP_i)]^{a_1} \partial_{P_4^{a_1}} +[S^{(2)}(\pP_i)]^{a_1a_2} \partial_{P_4^{a_1}}\partial_{P_4^{a_2}}+\ldots \Big)\delta(\sum_i \pP_i)
\end{align}
The function $S^{(k)}(\pP_i)$ has $k$ $SO(d,1)$ indices which contract with $k$ $\pX$'s. They give rise to $k$ derivatives on the momentum conserving delta function in the momentum space. Astute reader will notice that this decomposition is ambiguous. This is because homogeneity in $\pX$ is not well defined as illustrated by the example below.
\begin{equation}
    \int d\pX (\sum \pP_i) \cdot \pX\, e^{i(\sum \pP_i)\cdot X}=-i\int d\pX \, \pX\cdot \partial_{\pX} e^{i(\sum \pP_i)\cdot X}=i\int d\pX e^{i(\sum \pP_i)\cdot X}
\end{equation}
We fix this ambiguity by demanding $S^{(k)}(\pP_i)$ only to be function of $\pP_i$ for $i=1,2,3$ and \emph{not} of $\pP_4$. This can always be done by setting $\pP_4=\sum\pP_i-(\pP_1+\pP_2+\pP_3)$ and eliminating $\sum\pP_i$ by introducing $-i\partial_{\pX}$ acting on the exponential factor $e^{i(\sum_i\pP_i)\cdot \pX}$. 
At ${\cal O}(1/\tR^2)$, the only non-zero terms are  $S^{(k)}(\pP_i)$ for $k=0,\ldots,3$. They are
\begin{align}
    S^{(0)}(\pP_i)&=1+\frac{1}{\tR^2}\frac{12 + 13 d -d^{3}}{12 m^{2}}\notag\\
    S^{(1)}(\pP_i)^{a_1}&=\frac{1}{\tR^2}\frac{3(d+2)}{2 m^2} \big( \sum \limits_{i=1}^3 \pP_i^a \big)\notag\\
    S^{(2)}(\pP_i)^{a_1 a_2}&=\frac{1}{\tR^2}\Big(- \frac{d}{4 m^2} \big( \sum \limits_{i=1}^3 \pP_i^a \pP_i^b \big) + \frac{6 +  d }{4 m^{2}} \big( \sum \limits_{i=1}^3 \pP_i^a \big) \big( \sum \limits_{i=1}^3 \pP_i^b \big)  + \frac{1}{2} \eta^{a b} \Big)\notag\\
    S^{(3)}(\pP_i)^{a_1 a_2 a_3}&=\frac{1}{\tR^2}\frac{1}{6 m^2} \Big( - \big( \sum \limits_{i=1}^3 \pP_i^a \pP_i^b \pP_i^c \big) + \big( \sum \limits_{i=1}^3 \pP_i^a \big) \big( \sum \limits_{i=1}^3 \pP_i^b \big)\big( \sum \limits_{i=1}^3 \pP_i^c \big) \Big).
\end{align}
In particular this means that, in general, $S^{(0)}(\pP_i)$ can be taken as a function of Mandelstam variables $s\equiv -(\pP_1+\pP_2)^2, \,t\equiv -(\pP_1+\pP_3)^2$. We \emph{define} the AdS S-matrix to be this function $S(s,t)\equiv S^{(0)}(\pP_1,\pP_2,\pP_3)$. With this definition, the S-matrix for $\phi^4$ contact diagram is $1+\frac{1}{\tR^2}\frac{12 + 13 d -d^{3}}{12 m^{2}}+{\cal O}(1/\tR^4)$. Later, we will define the AdS S-matrix by stripping off the contribution coming the external leg corresponding to $\pP_4$. With this definition the leading order S-matrix agrees with the flat-space S-matrix as desired.

\subsection{Ward identity for $M_a$}\label{ward}
It might seem that our definition of the AdS S-matrix forgets a lot of information that is in the momentum space correlators viz. the terms containing derivatives of the delta function but that is not so. We will now show that the full momentum space correlator is fixed by the AdS S-matrix using the conformal Ward identities for generators $M_a$.

The action of $M_a$ in momentum space is computed in equation \eqref{Ma-action}. In order to imposing Ward identity $\sum_{i} M^{(i)}_{a} =0$ on the correlator $G(\pP_i)$ in perturbation theory, it is convenient to expand each $S^{(k)}(\pP_i)$ as,
\begin{equation}
[S^{(k)}(\pP_i)]^{a_1 \ldots a_k} = \sum_{n=0}^{\infty} \frac{1}{\tR^{2n}}[S^{(k,n)}(\pP_i)]^{a_1 \ldots a_k}.
\end{equation}
Note that we have written $S$ as an expansion in $1/\tR^2$ and not $1/\tR$ because the propagator and vertices only contain even powers of $1/\tR$. Also the generator $M_a$ has an expansion in $1/\tR^2$. 
Now we are ready to implement the $M_a$ Ward identity. At $\CO(1)$,
\begin{equation}
\sum_{i=1}^4 \pP_{i,a}  \Big[ \sum_{k} [S^{(k,0)}(\pP_{i})]^{a_1 \ldots a_k} \partial_{\pP_{4}^{a_1}} \dots \partial_{\pP_{4}^{a_k}} \ \delta \big( \sum \pP_{i} \big) \Big] =0 
\end{equation}
This is solved by $[S^{(k,0)}(\pP_{i})]^{a_1 \ldots a_k}=0 \ \forall \ k \geq 1$ and any function $S^{(0,0)}(\pP_{i})$, since $\sum_{i}\pP_{i}^{a}$ is killed by $\delta \big( \sum \pP_{i} \big)$ but not its derivatives. 

At $\CO(1/ \tR^2)$, we have,
\begin{align}
\nonumber
& \sum_{i=1}^4 \pP_{i,a}  \Big[ \sum_{k} [S^{(k,1)}(\pP_{i})]^{a_1 \ldots a_k} \partial_{\pP_{4}^{a_1}} \dots \partial_{\pP_{4}^{a_k}} \ \delta \big( \sum \pP_{i} \big) \Big]  + \sum_{i}  M_{i,a}^{(1)} \Big[ S^{(0,0)}(\pP_{i})  \ \delta \big( \sum \pP_{i} \big) \Big] =0 \\
\nonumber
\Rightarrow &  \sum_{i} \pP_{i,a}  \Big[ \sum_{k} [S^{(k,1)}(\pP_{i})]^{a_1 \ldots a_k} \partial_{\pP_{4}^{a_1}} \dots \partial_{\pP_{4}^{a_k}} \ \delta \big( \sum \pP_{i} \big) \Big] \\
  = &-\sum_{i}  \frac{\pP_{i,a}}{4 m_{i}^2} (d - 2d \,\,\pP_i \cdot \partial_{\pP_{i}} -2 \pP^{a}_{i} \pP^{b}_{i} \partial_{\pP^{a}_{i}}\partial_{\pP^{b}_{i}} -2m^2 \partial_{\pP_{i}} \cdot \partial_{\pP_{i}} ) \Big[ S^{(0,0)}(\pP_{i})  \ \delta \big( \sum \pP_{i} \big) \Big].
\end{align}
The $k$-th term on the left hand side seems to have $k$ derivatives acting on the delta function, however it instead has only $k-1$ derivatives acting on the delta function. This is best seen by integrating the left hand side over $\pP_4$  with respect to a test function, say $f(\pP_4)$ using partial integration. The term with $f^{(k)}(\pP_4)$ vanishes due to $\sum_{i=1}^4\pP_i=0$. However, there is another term with $f^{(k-1)}(\pP_4)$ obtained by one of the $\partial_{\pP_4}$ acting on $\pP_4$ that appears in $\sum_{i=1}^4\pP_i$. The right hand side has terms with $0,1,2$ derivatives acting on the test function. This means $S^{(k,1)}=0$ for $k>3$. We solve this equation for $S^{(k,1)}$ for $k=1,2,3$ in terms of the leading AdS S-matrix $S^{(0,0)}(\pP_i)$ and get,
\begin{align}\label{ward-solve}
[S^{(1,1)}(\pP_{i})]^{a_1} &=\Big[- \frac{d}{2} \sum_{j=1}^{3} \frac{\pP_{j}^{a_{1}}\pP_{j}^{b}}{m_{j}^{2}} \partial_{\pP_{j}^{b}} - \frac{1}{2} \sum_{j=1}^{3} \frac{\pP_{j}^{a_{1}} \pP_{j}^{b} \pP_{j}^{c} }{m_{j}^{2}} \partial_{\pP_{j}^{b}} \partial_{\pP_{j}^{c}} - \frac{1}{2}  \sum_{j=1}^{3} \pP_{j}^{a_{1}} \partial_{\pP_{j}}^{b}\partial_{\pP_{j},b} \notag\\
& +\frac{d}{4} \sum_{j=1}^{3} \frac{\pP_{j}^{a_{1}} }{m_{j}^{2}}   +\frac{5d+12}{4 m_{4}^{2}} \big( \sum \limits_{i=1}^3 \pP_i^{a_{1}} \big)  \Big]  \ S^{(0,0)}(\pP_{i}) \notag \\
[S^{(2,1)}(\pP_{i})]^{a_1 a_2} &= \Big[ \frac{1}{2} \eta^{a_{1} a_{2}} - \frac{1}{2} \sum_{j=1}^{3} \pP_{j}^{a_{1}} \partial_{\pP_{j}}^{a_{2}} - \frac{1}{2} \sum_{j=1}^{3} \frac{\pP_{j}^{a_{1}} \pP_{j}^{a_{2}} \pP_{j}^{b} }{m_{j}^{2}} \partial_{\pP_{j}^{b}} + \frac{d+6}{4 m_{4}^{2}} \big( \sum \limits_{i=1}^3 \pP_i^{a_{3}} \big) \big( \sum \limits_{i=1}^3 \pP_i^{a_{2}} \big) \notag\\
& -\frac{d}{4} \sum_{j=1}^{3} \frac{\pP_{j}^{a_{1}} \pP_{j}^{a_{2}} }{m_{j}^{2}}    \Big]  \ S^{(0,0)}(\pP_{i}) \notag\\
[S^{(3,1)}(\pP_{i})]^{a_1a_2a_3} &=\Big[- \frac{1}{6}\sum_{j=1}^{3} \frac{\pP_{j}^{a_{1}} \pP_{j}^{a_{2}} \pP_{j}^{a_{3}} }{m_{j}^{2}}  + \frac{1}{6 m_{4}^{2}} \big( \sum \limits_{i=1}^3 \pP_i^{a_{1}} \big) \big( \sum \limits_{i=1}^3 \pP_i^{a_{2}} \big)\big( \sum \limits_{i=1}^3 \pP_i^{a_{3}} \big) \Big] \ S^{(0,0)}(\pP_{i}).
\end{align}
For the case of contact diagrams, these expressions precisely match with those derived from direct computation of the integral \eqref{contact1}.

In the same way, at $1/\tR^4$ order, $S^{(k,2)}(\pP_i)$ for $k\geq 1$, are fixed in terms of $S^{(0,0)}$ and $S^{(0,1)}$ which are determined above. Moreover, as the corresponding right hand side has finitely many $\pP_4$ derivatives only first few  $S^{(k,2)}$'s are non-zero.
In this way, $S^{(k)}(\pP_i)$ are completely determined from $S^{(0)}(\pP_i)$ in $1/\tR^2$ perturbation theory. Let us emphasize that the complete perturbative expansion of $S^{(0)}$ i.e. all the functions $S^{(0,m)}$ serve as independent data for this procedure.
In other words, all the information in momentum space correlator $\hat G(\pP_{i})$ is stored in AdS S-Matrix $S^{(0)}(\pP_i)$. 

\subsubsection{Derivative contact diagram}
Let us now present the computation of a slightly more complicated diagram: $\nabla^a \phi\nabla_a\phi\,\phi\phi$ in the s-channel as shown in the diagram. In terms of the local coordinates $\pX$ on AdS, the covariant derivative and the metric are
\begin{align}\label{metric}
    \nabla_a&=\partial_a-\Gamma_a,\quad \Gamma_{ab}^c=-\frac{1}{\tR^2} \pX^c g_{ab}(\pX),\notag\\
    g_{ab}(\pX)&=\eta_{ab}-\frac{\eta_{am}\eta_{bn}\pX^m\pX^n}{\tR^2+\eta_{cd}\pX^c\pX^d}.
\end{align}
For scalars we simply use $\nabla_a \phi=\partial_a\phi$.
With this, the contact diagram is
\begin{align}
    G(\pP_i)\prod_i |\pP_i|^{d-\Delta_i}&=\int \frac{\tR\, d\pX}{\sqrt{\tR^2-\pX^2}} \prod_{i=3}^4 g_{\partial B}\Big(\frac{\pP_i\cdot \pX}{m_i}\Big)e^{i(\pP_3+\pP_4)\cdot \pX}  \notag\\
    &\times \nabla^a \Big[g_{\partial B}\Big(\frac{\pP_1\cdot \pX}{m_1}\Big)e^{i\pP_1\cdot \pX}\Big]\nabla_a \Big[g_{\partial B}\Big(\frac{\pP_2\cdot \pX}{m_2}\Big)e^{i\pP_2\cdot \pX}\Big]\\
    &\equiv\int  d\pX f(\pP_i,\pX)e^{i\sum_{i=1}^{4}\pP_i\cdot\pX}.
\end{align}
The last line is the definition of the function $f(\pP_i,\pX)$. To compute the AdS S-matrix,   we first replace $\pP_4$ in $f$ by $\pP_4=\sum_{i=1}^{4} \pP_i-(\pP_1+\pP_2+\pP_3)$ and substitute $\sum_{i}\pP_i$ by $-i\partial_{\pX}$. The derivative is then partially integrated to act on the other terms in $f$. After doing this, $\pP_4$ doesn't appear in this factor. By abuse of notation let us denote this factor by $f(\pP_i,\pX)$ but now $i$ only takes the values $i=1,2,3$. The AdS S-matrix is then easily computed, simply by setting $\pX=0$ in $f(\pP_i,\pX)$. This is because  $\pX$ written as  $-i\partial_{\pP_4}$ and because $f(\pP_i,\pX)$ doesn't contain $\pP_4$, this derivative only acts on the momentum-conserving-$\delta$-function. The AdS S-matrix is precisely the piece having no derivative on the $\delta$-function. Hence it is obtained by simply setting $\pX=0$. At leading order the AdS S-matrix agrees with the flat-space S-matrix as expected. We get,
\begin{align}
    S^{(0)}(\pP_i)&=\Big(m^2-\frac{s}{2}\Big)\Big(-1+\frac{1}{\tR^2}\frac{-12 - 19 d +d^{3}}{12 m_{\phi}^{2}}\Big)\\
    S^{(1)}(\pP_i)&=\frac{1}{\tR^2}\Big(m^2-\frac{s}{2}\Big)\Big( -\frac{d+3}{ m_{\phi}^2} (\pP_{1}^{a}+\pP_{2}^{a}) - \frac{3(d+2)}{2 m_{\phi}^2} \pP_{3}^{a} \Big)\notag\\
    S^{(2)}(\pP_i)&=\frac{1}{\tR^2}\Big(m^2-\frac{s}{2}\Big)\Big( \frac{d+2}{4 m_{\phi}^2} (\pP_1^a \pP_1^b + \pP_2^a \pP_2^b)+ \frac{d}{4 m_{\phi}^2} \pP_3^a \pP_3^b - \frac{6 +  d }{4 m_{\phi}^{2}} \big( \sum \limits_{i=1}^3 \pP_i^a \big) \big( \sum \limits_{i=1}^3 \pP_i^b \big)  - \frac{1}{2} \eta^{a b} + \frac{1}{\pP_{12}} \pP_1^a \pP_2^b \Big)\notag\\
    S^{(3)}(\pP_i)&=\frac{1}{\tR^2}\Big(m^2-\frac{s}{2}\Big)\frac{1}{6 m_{\phi}^2}\Big( \big( \sum \limits_{i=1}^3 \pP_i^a \pP_i^b \pP_i^c \big) - \big( \sum \limits_{i=1}^3 \pP_i^a \big) \big( \sum \limits_{i=1}^3 \pP_i^b \big)\big( \sum \limits_{i=1}^3 \pP_i^c \big) \Big).\notag
\end{align}
Again, this agrees with the general relation obtained from the conformal Ward identity \eqref{ward-solve}.
This prescription is easily extended to compute general contact diagrams. In section \ref{feynman} we give Feynman-like rules to compute general Witten diagrams in $1/\tR^2$ perturbation theory.

\subsection{Bulk to bulk propagator}\label{propa}
Another crucial ingredient required for computing Witten diagrams is the bulk-to-bulk propagator $G_{BB}(\pY,\pX)$ of a particle of mass $m_\xi$. In this section we will compute the Fourier transform with respect to one of its points $\pX$ to get $\hat G_{BB}(\pP;\pX)$. Instead of directly Fourier transforming the position space expression for the propagator, we will do this by solving the differential equation.

\begin{align}\label{green's}
\big[-\nabla_{\pX,a} \nabla^{a}_{\pX} + m_\xi^{2} \big] \ G_{BB}(\pY,\pX) &= \frac{1}{\sqrt{g(\pX)}} \delta(\pY-\pX) \notag\\
\big[-\nabla_{\pX,a} \nabla^{a}_{\pX} + m_\xi^{2} \big] \ \hat G_{BB}(\pP;\pX) &= \frac{1}{\sqrt{g(\pX)}} e^{i\pP\cdot \pX} 
\end{align}
In the second line we have Fourier transformed the point $\pY$ to momentum space $\pP$. The AdS metric is given in equation \eqref{metric}. Using this, the equation becomes
\begin{equation}\label{green's_2}
\big[ -\partial_{\pX} \cdot \partial_{\pX} + m_\xi^2 - \frac{1}{\tR^2} ( \pX^{a} \pX \cdot \partial_{\pX} \partial_{\pX,a} + (d+1) \pX \cdot \partial_{\pX} ) \big] \ \hat G_{BB}(\pP;\pX) = \Big( 1+ \frac{\pX \cdot \pX}{\tR^2} \Big)^{\frac{1}{2}} e^{i\pP\cdot \pX} 
\end{equation}
where, $(\cdot)$ is with respect to $\eta_{ab}$. Expanding $\hat G_{BB}(\pP;\pX)$ in $1/\tR^2$ as $\sum_n \tR^{-2n}\hat G_{BB}^{(n)}(\pP;\pX)$, we get
\begin{align}\label{BB-prop}
    \hat G_{BB}^{(0)}(\pP;\pX) &= \frac{e^{i \pP \cdot \pX}}{\pP  \cdot \pP + m_\xi^2}\notag\\
    \hat G_{BB}^{(1)}(\pP;\pX) &= \frac{e^{i \pP \cdot \pX}}{\pP  \cdot \pP + m_\xi^2} \Big(  \frac{\pX \cdot \pX}{2} + \frac{-(\pP \cdot \pX)^2 + i(d-1)\pP \cdot \pX}{(\pP \cdot \pP + m_\xi^2)}\notag\\
    &+ \frac{4 i m_\xi^2 \pP \cdot \pX - (d-1)\pP \cdot \pP}{(\pP \cdot \pP + m_\xi^2)^2} +\frac{(d-7) m_\xi^2 \pP \cdot \pP + (d+1)m_\xi^4}{(\pP \cdot \pP + m_\xi^2)^3} \Big).
\end{align}
Higher order corrections $\hat G_{BB}^{(i)}(\pP;\pX), i\geq 2$ can be computed straightforwardly. Although redundant, it turns out to be useful to think of  $\hat G_{BB}(\pP;\pX)$ as a function of $SO(d+1,1)$ invariant quantities as $(\pX^2, \pP\cdot \pX,(\pP^2+m_\xi^2)^{-1})$ so that $\hat G_{BB}$ can be written as a differential operator acting on $\hat G_{BB}^{(0)}(\pP;\pX)$ i.e.
\begin{align}
    \hat G_{BB}^{(0)}(\pP;\pX)=g_{BB}(-\partial^2_\pP, -i\pP\cdot \partial_{\pP},(\pP^2+m_\xi^2)^{-1})\,\frac{e^{i \pP \cdot \pX}}{\pP  \cdot \pP + m_\xi^2}
\end{align}

\section{Feynman Rules in AdS}\label{feynman}
Witten diagrams are used to compute correlation functions in a boundary conformal field theory that corresponds to weakly coupled bulk dual. Due to lack of translational symmetry in AdS, these diagrams need to be computed in position space which makes the calculation of higher loop diagrams very cumbersome. In this section, we will develop Feynman-like rules to compute Witten diagrams in $\eP$-space in $1/\tR^2$ perturbation theory. They have the property that they reduce to ordinary Feynman rules in the strict $\tR\to \infty$ limit.

\subsection{Vertex}
\begin{figure}[h]
    \begin{center}
    \includegraphics[scale=1]{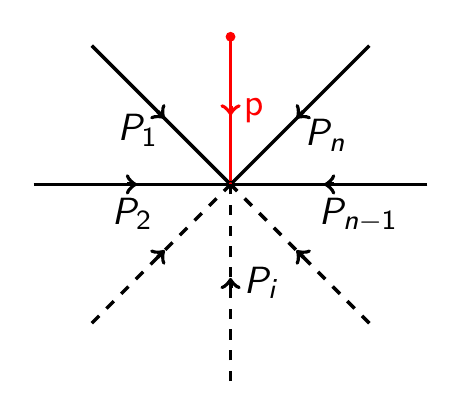}
    \end{center}
    \caption{An $n$-point derivative vertex. We have denoted the momentum injection $\sfp$ in red.}
    \label{contact_rule}
\end{figure}

Although, Feynman rules can be written down just as easily for an $n$-point vertex for general $n$, we will consider the case of $n=4$ for simplicity. A general $4$-pt vertex can be written as $V_{i,j}\equiv \nabla^{i+j}\phi_1 \nabla^i \phi_2 \nabla^j \phi_3\phi_4$. Although the fields $\phi_i$ are identical, we have assigned them a subscript for convenience. Here the indices of the covariant derivatives acting on $\phi_2$ and $\phi_3$ are contracted with the same acting on $\phi_1$ using the metric. Explicit form of the covariant derivative and the metric in $\pX$ coordinates is given in equation \eqref{metric}. Schematically the vertex $V_{i,j}(\pZ)$ in position-space takes the form
\begin{align}
    V_{i,j}(\pZ)=g(\pZ)^{i+j}D_{i+j}(\pZ,\partial_{\pZ})\phi_1(\pZ) D_{i}(\pZ,\partial_{\pZ}) \phi_2(\pZ) D_{j}(\pZ,\partial_{\pZ}) \phi_3(\pZ)\phi_4(\pZ).
\end{align}
where $D_i^{a_1a_2,\ldots,a_i}(\pZ,\partial_{\pZ})$ is the covariant derivative operator $\nabla^i$ with open $SO(d,1)$ indices $a_1,\ldots,a_i$. We take them to be ``normal-ordered'' i.e. 
$\partial_Z$'s acts on $\phi(Z)$ before multiplying by $Z$'s. For example,
\begin{align}
    \nabla_a \, \phi(\pZ)&= \partial_a \, \phi(\pZ) \notag \\
    \nabla_a \nabla_b \, \phi(\pZ)&= \partial_a\partial_b \, \phi(\pZ)-\Gamma_{ab}^c(\pZ)\, \partial_c\,\phi(\pZ) \qquad \ldots {\rm etc.}
\end{align}
Also, $g(\pZ)^{i+j}$ denotes $i+j$ factors of the metric used to contract indices of covariant derivatives acting on $\phi_2$ and $\phi_3$ with those acting on $\phi_1$.
To compute this vertex in $\pP$-space, we substitute the wavefunction $\phi_i(Z)=e^{i\pP_i\cdot\pZ}$.
\begin{align}
    V_{i,j}(\pP_i)&=\int d\pZ \sqrt{g(\pZ)}g(\pZ)^{i+j} D_{i+j}(\pZ,\partial_{\pZ})e^{i\pP_1\cdot\pZ} D_{i}(\pZ,\partial_{\pZ}) e^{i\pP_2\cdot\pZ}(\pZ) D_{j}(\pZ,\partial_{\pZ}) e^{i\pP_3\cdot\pZ}e^{i\pP_4\cdot\pZ}\notag\\
    &=\sqrt{g(-i\partial_{\pP_4})} g(-i\partial_{\pP_4})^{i+j}D_{i+j}(-i\partial_{\pP_4},i\pP_1) D_{i}(-i\partial_{\pP_4},i\pP_2)  D_{j}(-i\partial_{\pP_4},i\pP_3) \delta(\sum_{i=1}^4 \pP_i)\notag\\
    &=\sqrt{g(-i\partial_{\sfp})} g(-i\partial_{\sfp})^{i+j}D_{i+j}(-i\partial_{\sfp},i\pP_1) D_{i}(-i\partial_{\sfp},i\pP_2)  D_{j}(-i\partial_{\sfp},i\pP_3) \delta(\sfp+\sum_{i=1}^4 \pP_i)|_{\sfp=0}.
\end{align}
In the second line, we have done $\pZ$ integral to get the momentum conserving $\delta$ function. The explicit factors of $\pZ$ are written as $-i\partial_{\pP_4}$ using $-i\partial_{\pP_4}e^{i\pP_4\cdot\pZ}=\pZ e^{i\pP_4\cdot\pZ}$. We could have used any $\pP_i$ but we used $\pP_4$ so that it only acts on the $\delta$ function and on no other factor. For future convenience, it is useful to introduce a new  ``momentum'' variable $\sfp$ that is injected into the vertex and the derivative is taken with respect to this injected momentum and not $\pP_4$. This is what we have done in the third line. Graphically an $n$-point vertex is presented as in figure \ref{contact_rule}. 

It is important to keep track of terms containing derivatives of the $\delta$-function in the vertex factor. This because the vertex factor by itself doesn't constitute the full Witten diagram but is only its part. Remember, the AdS S-matrix is computed by discarding the derivatives of the $\delta$-function but only of the whole $\pP$-space Witten diagram.

\subsection{Propagators}
The $\pP$-space treatment of propagators takes a similar  route. The propagator with one of its end in $\pP$-space and the other in position-space takes the form 
\begin{align}
    G(\pP;\pX')=f(\pP;\pX') e^{i\pP\cdot \pX'}
\end{align}
Here $G(\pP;\pX')$ can either be a boundary-to-bulk or bulk-to-bulk propagator. In both cases, $\pX'$ is taken to be the point in the bulk. The function $f(\pP;\pX')$ is $1$ and $1/(\pP^2+m^2)$ at leading order for boundary to bulk and bulk to bulk propagators respectively but contains non-trivial $\pX'$ dependance at sub-leading orders. To take the point $\pX'$ to momentum-space, we simply Fourier transform
\begin{align}\label{g-inject}
    G(\pP,\pP')&=\int d\pX'  \,f(\pP;\pX') e^{i(\pP+\pP')\cdot \pX'}\notag\\
    &= \,f(\pP;-i\partial_\sfp) \delta(\pP+\pP'+\sfp)|_{\sfp=0}.
\end{align}
\begin{figure}[h]
    \begin{center}
    \includegraphics[scale=1]{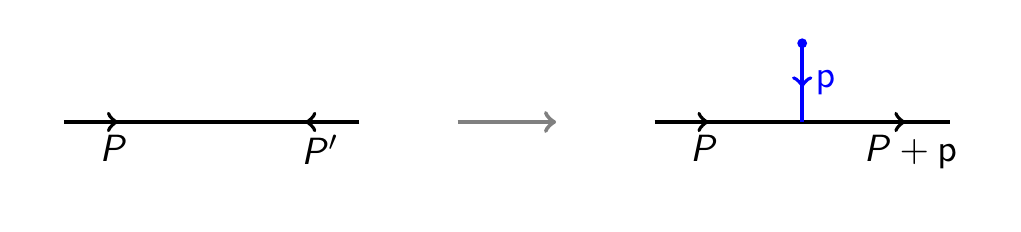}
    \end{center}
    \caption{Bulk to bulk propagator. We have denoted the momentum injection $\sfp$ in blue.}
    \label{prop_diagram}
\end{figure}
In the second line, we have written the propagator with the injection of auxiliary AdS momentum $\sfp$ to account for momentum non-conservation. It is represented graphically as in figure \ref{prop_diagram}. The function $f(\pP;\pX')$ is $f_{\partial B}(\pP;\pX')$ and $f_{BB}(\pP;\pX')$ for boundary to bulk and bulk to bulk propagators respectively
\begin{align}
    f_{\partial B}(\pP;\pX')|\pP|^{d-\Delta}&=g_{\partial B}(\frac{\pP\cdot \pX'}{m}) \notag\\
    f_{B B}(\pP;\pX')&=g_{BB}(\pX'^2, \pP\cdot \pX',(\pP^2+m_\xi^2)^{-1}).
\end{align}
The function $g_{\partial B}$ and $g_{BB}$ are defined in equations \eqref{leg} and \eqref{BB-prop} respectively. Due to this specific form of the function $f_{\partial B}$ the boundary to bulk propagator can be simplified as
\begin{equation}\label{g1-simple}
    G_{\partial B}(\pP,\pP')=m^{\Delta-d}g_{\partial B}\Big(\frac{-i\partial_\sigma}{m}\Big)\delta(\sigma \pP+\pP')|_{\sigma=1}.
\end{equation}
If we do this for the fourth leg with momentum $\pP_4$, we will get an S-matrix that is explicitly a function of $\pP_4$. As emphasized earlier, we would like to swap the $\pP_4$ dependence for $-\sum_{i=1}^{3}\pP_i$. Recall that this is not trivial due to presence of derivatives of the momentum conserving delta function in the correlator. We will deal with it by isolating only the piece of $G_{\partial B}(\pP_4,\pP')$ that is proportional to the delta function. To that effect, let us integrate the propagator with respect to a test function of $\pP_4$ using partial integration.
\begin{align}
    \int d\pP_4\, \, f(\pP_4)\,\,G_{\partial B}(\pP_4,\pP')=\int d\pP_4\, \,f(\pP_4)\, m_4^{\Delta-d}g_{\partial B}\Big(\frac{-i}{m_4}\pP_4\cdot \partial_{\pP_4}\Big) \delta(\pP_4+\pP')
\end{align}
Keeping only the terms with $f(-\pP')$ and not the derivatives, we get the following factor for the $\pP_4$ leg to ${\cal O}(1/\tR^2)$.
\begin{align}
    G_{\partial B}(\pP_4,\pP')|_{\delta}=1+ \frac{1}{\tR^{2}} \Big(\frac{12+13 d-d^3}{12 m_4^{2}}\Big)
\end{align}
This factor is independent of $\pP'$ to all orders because, the number of derivatives $\partial_{\pP_4}$ annihilate all the $\pP_4$'s when they are not acting on the test function. So it is convenient to strip off this factor while defining the S-matrix. From now on, we will attach only the factor of $1$ to the $\pP_4$ leg.

The bulk to bulk propagator can also be simplified
\begin{align}\label{g2-simple}
    G_{BB}(\pP,\pP')=\tilde g_{BB}(-\partial_\sfp^2,-i\pP\cdot \partial_\sfp, \partial_\alpha) \frac{\delta(\pP+\pP'+\sfp)}{\pP^2+\alpha m_\xi^2}|_{(\alpha,\sfp)=(1,\vec{0})}.
\end{align}
Here the function $\tilde g_{BB}(x,y,z)$ is obtained from $g_{BB}(x,y,z)$ by replacing $z^{n}\to(-z/m_\xi^2)^n/n! $.
Using the form of propagators in equations \eqref{g1-simple} and \eqref{g2-simple}, allows one to compute the Witten diagram - even the higher loop ones -  as an almost standard Feynman diagram but with more parameters i.e. set of $(\sigma, \alpha, \beta, \sfp)$ and then acting on it by appropriate differential operators in these parameters.

\subsection{Diagram computation}
After describing the computation of the vertex and propagators we are now in a position to formulate Feynman-like rules to compute Witten diagrams. Here we give the rules for four-point AdS S-matrix but they are easily generalized to higher point S-matrix.

\begin{itemize}
    \item Decorate the Witten diagram with momentum injections at each vertex and at every internal propagator and replace external momenta $\pP_i\to \sigma_i\pP_i$ for $i=1,2,3$.
    \item Conserve the momenta at every vertex (including the tri-valent vertex corresponding to the propagator) and solve all the internal momenta and $\pP_4$ in terms of external momenta $\pP_1,\pP_2, \pP_3$, injected momenta and loop momenta. 
    \item Associate the following factors for each vertex, internal propagator and external legs.
    \begin{enumerate}
        \item Vertex:
            \begin{figure}[H]
            \begin{center}
            \includegraphics[scale=1]{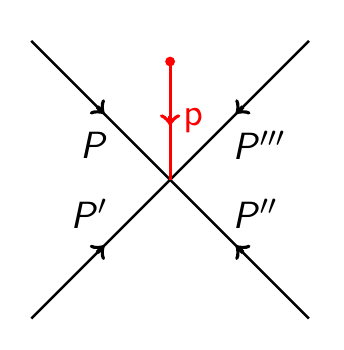}
        \end{center}
            \qquad 
            \[\sqrt{g(-i\partial_{\sfp})} g(-i\partial_{\sfp})^{i+j}D_{i+j}(-i\partial_{\sfp},i\pP_1) D_{i}(-i\partial_{\sfp},i\pP')  D_{j}(-i\partial_{\sfp},i\pP'')\]
            
        \end{figure}
        Similarly for higher point vertices.
        \item Internal propagator:
            \begin{figure}[H]
            \begin{center}
            \includegraphics[scale=1]{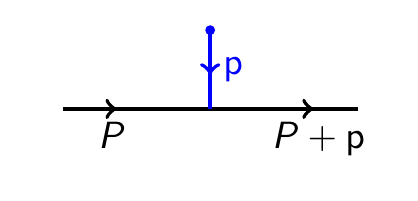}
        \end{center}
            \qquad \\
            \[\tilde g_{BB}(-\partial_\sfp^2,-i\pP\cdot \partial_\sfp, \partial_\alpha) \frac{1}{\pP^2+\alpha m_\xi^2}\]
            \end{figure}
        \item External leg:
            \begin{figure}[H]
            \begin{center}
            \includegraphics[scale=1]{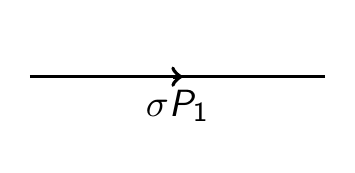}
        \end{center}
            \qquad \\
            \[g_{\partial B}\Big(\frac{-i\partial_\sigma}{m}\Big)\]
           
            \end{figure}
        Here we have dropped the constant, unimportant factor $m^{\Delta-d}$. In the leg corresponding to the external momentum $\pP_4$, we simply associate the factor $1$.
    \end{enumerate}
    \item Truncate the resulting quantity to desired order in $1/\tR^2$. 
    Perform loop integrals and act with all the differential operators and finally set all the $\alpha$'s and $\sigma$'s to $1$ and momentum injections to $0$.
\end{itemize}
With these modified rules at hand we can now easily compute Witten diagrams in $1/\tR$ perturbation theory.  In the rest of the section we will apply these Feynman rules to compute scalar exchange diagram and a scalar bubble diagram.

\subsection{Scalar exchange diagram}
\begin{figure}[t]
    \begin{center}
    \includegraphics[scale=0.5]{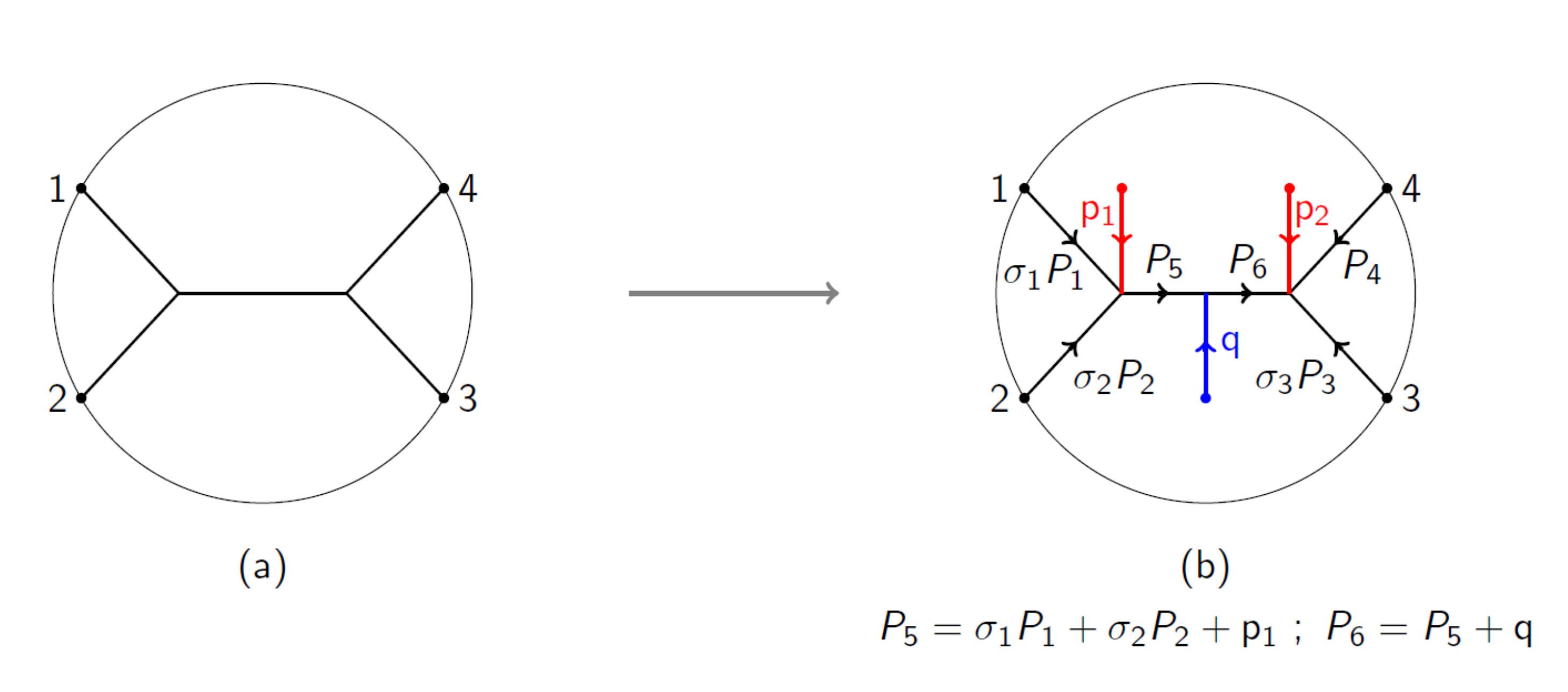}
    \end{center}
    \caption{Exchange diagram with momentum injections at the two vertices and at the internal propagator. All the internal momenta have been solved in terms of external momenta and momentum injections.}
    \label{exchange_dia}
\end{figure}

In figure \ref{exchange_dia}, we have an exchange Witten diagram for external scalars of mass $m$ and exchanged scalar of mass $m_\xi$. The internal momenta and $\pP_4$ are solved in terms of $\pP_1,\pP_2,\pP_3$ and momentum injections. Following the above rules, we get

\begin{align}
    D_{exch}
    &=\prod_{i=1}^3\, g_{\partial B}\Big(\frac{-i \partial_{\sigma_i}}{m}\Big) \,\sqrt{g(-i\partial_{\sfp_1})} \sqrt{g(-i\partial_{\sfp_2})} \,\,\tilde g_{BB}(-\partial_\sq^2,-i(\sigma_1\pP_1+\sigma_2\pP_2+\sfp_1)\cdot \partial_\sq, \partial_\alpha)\notag\\
    &\frac{1}{(\sigma_1\pP_1+\sigma_2\pP_2+\sfp_1)^2+\alpha m_\xi^2}|_{\alpha=1,\sigma_i=1,\sfp_i=\sq=0}\\
    &=\prod_{i=1}^2\, g_{\partial B}\Big(\frac{-i \partial_{\sigma_i}}{m}\Big) \,\tilde g_{BB}(0,0, \partial_\alpha)\sqrt{g(-i\partial_{\sfp_1})} \frac{1}{(\sigma_1\pP_1+\sigma_2\pP_2+\sfp_1)^2+\alpha m_\xi^2}|_{\alpha=1,\sigma_i=1,\sfp_1=0}\notag\\
    &=\frac{-1}{s- m_{\xi}^2}+\frac{1}{\tR^2}\Big( \frac{2 m_{\xi}^{4}(4 m^{2} - m_{\xi}^{2})}{m^{2} \ (s-m_{\xi}^{2})^{4}} +  \frac{ (d-6) m_{\xi}^{4}-2 (d-8) m^{2} m_{\xi}^{2} }{m^{2} \ (s-m_{\xi}^{2})^{3}}\notag\\
    & +  \frac{(16-6 d) m^{2}+(5 d-12) m_{\xi}^{2}}{2 m^{2} \ (s-m_{\xi}^{2})^{2}} + \frac{3 d-4}{2 m^{2} (s-m_{\xi}^{2})}  \Big) \notag
\end{align}
In the first equality we have the factors coming from the three external legs, vertices and the internal propagator respectively. Unlike the usual Feynman rules, these factors are not c-numbers but rather involve derivatives with respect to the external and injected momenta. In the second equality, the expression is simplified considerably because derivatives in some of the differential operators trivially evaluate to zero. Finally, using the explicit expression for $g_{\partial B}$ and $g_{BB}$ we can readily evaluate the diagram and the answer is given in the third equality. 

\subsection{Bubble diagram}
\begin{figure}[t]
    \begin{center}
    \includegraphics[scale=0.5]{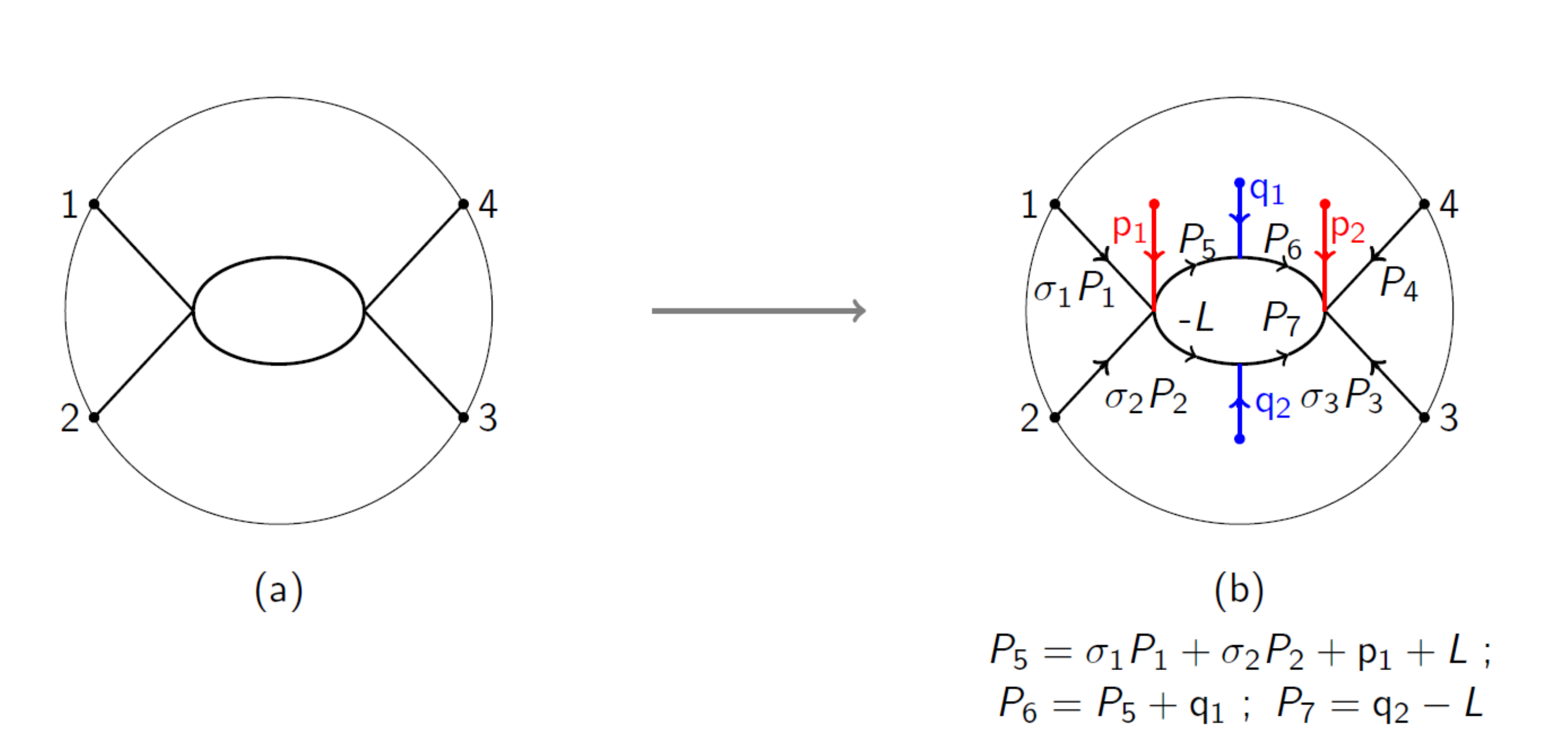}
    \end{center}
    \caption{A bubble diagram with all the momentum injections.}
    \label{bubble_dia}
\end{figure}

Now we move on to the scalar bubble diagram in $\phi^4$ theory  as shown in figure \ref{bubble_dia}. We have solved for all the internal momenta and $\pP_4$ in terms of the external momenta, injected momenta $\sq_{1,2}$ and loop momentum $\pL$. This diagram has been first computed in \cite{Fitzpatrick:2011hu} using harmonic analysis to express product of two bulk-bulk propagators in terms of sum of bulk-bulk propagators. More recently, this diagram played a prominent role in \cite{Aharony:2016dwx} in the context of understanding unitarity in AdS. A direct computation of this diagram is a bit cumbersome but as we will see below, computing it in $1/\tR$ perturbation theory using our Feynman rules is straightforward. Collecting all the factors we get
\begin{align}
    D_{bbl}=&\prod_{i=1}^2\, g_{\partial B}\Big(\frac{-i \partial_{\sigma_i}}{m}\Big) \,\tilde g_{BB}(0,0, \partial_{\alpha_1})\tilde g_{BB}(0,0, \partial_{\alpha_2})\sqrt{g(-i\partial_{\sfp_1})} \notag\\
    &\int d\pL_1 \,\,\frac{1}{(\sigma_1\pP_1+\sigma_2\pP_2+\sfp_1-\pL)^2+\alpha_1 m^2}\frac{1}{\pL^2+\alpha_2 m^2}
\end{align}
Here we have only kept the terms involving those differential operators that are non-trivial. These factors are the factors corresponding to the external legs $1,2$, the two bulk to bulk propagators and the vertex factor corresponding to only the first vertex (the one that couples to external legs $1,2$ and not $3,4$). As expected, the diagram also involves a loop integral. The loop integral can be done as in the usual Feynman diagrams. Even though the masses of the two virtual particles are the same, this loop integral appears in computing the flat space bubble diagram with virtual particles of different masses, $m\sqrt {\alpha_1}$ and $m\sqrt{\alpha_2}$.  Then we apply the appropriate differential operators to obtain the AdS S-matrix to $1/\tR^2$ order. 
\begin{align}
    D_{bbl}&=\int_0^1\, dx\, \frac{\Gamma(\frac{d-3}{2})}{(4\pi)^{\frac{d+1}{2}}\tilde s(x)^{\frac{3-d}{2}}} \\
    &+\frac{1}{\tR^2}\int_0^1\, dx\,\Big(\frac{ m^4 (2x-1)^2 \Gamma \left(\frac{9-d}{2} \right)}{ (4\pi) ^{\frac{d+1}{2}} \tilde{s}(x)^{\frac{9-d}{2}} }  +  \frac{ m^2 (4 (d-2) (x-1) x-1) \Gamma \left(\frac{7-d}{2} \right)}{ 2 (4\pi) ^{\frac{d+1}{2}} \tilde{s}(x)^{\frac{7-d}{2}} }  \notag\\
    &-  \frac{  (4 (9-4 d) (x-1) x + d (3 d-20)+31) \Gamma \left(\frac{5-d}{2} \right)}{ 4 (4\pi) ^{\frac{d+1}{2}} \tilde{s}(x)^{\frac{5-d}{2}} }  -  \frac{  (7-d)(3-d) (4 d-5) \Gamma \left(\frac{3-d}{2} \right)}{ 24 (4\pi) ^{\frac{d+1}{2}} \tilde{s}(x)^{\frac{3-d}{2}} } \Big)\notag.
\end{align}
where $\tilde s(x)=-x(1-x)s+m^2$ and $s=-(\pP_1+\pP_2)^2$ as usual. From the appearance of the specific Gamma functions in the $1/\tR^2$ terms, it is clear that the loop integral at $1/\tR^2$ order is either equally or more convergent in the UV compared to the flat space i.e. ${\cal O}(1)$ loop integral. It would be useful to prove a general result of this nature which states that the higher $1/\tR^2$ corrections are always either equally well-behaved or better behaved in the UV compared to their lower order counterparts.

\section{Connection with the Mellin Amplitude}\label{mellin}

In this section we will develop the dictionary between the $AdS$ S-matrix  and the mellin amplitude in $1/ \tR$ perturbation theory. 
\begin{align}
    G(\eW_i)&=(\eW_1\cdot\eW_4)^{-\Delta}(\eW_2\cdot\eW_3)^{-\Delta}\int  dS\, \int\,dT\, \Big(\frac{v}{u}\Big)^S\, v^T \,  M(\frac{S}{\tR},\frac{T}{\tR})\,\Gamma^2(S)\Gamma^2(T)\Gamma^2(\Delta-S-T) \notag\\
    {\rm where}\quad u&=\frac{(\eW_1\cdot \eW_2)(\eW_3\cdot \eW_4)}{(\eW_1\cdot \eW_3)(\eW_2\cdot \eW_4)},\quad v=\frac{(\eW_1\cdot \eW_4)(\eW_2\cdot \eW_3)}{(\eW_1\cdot \eW_3)(\eW_2\cdot \eW_4)}.
\end{align}
Here we have labeled the Mellin variables as $(S,T)$ to distinguish them from the ``Mandelstam" variables $(s\equiv -(\pP_1+\pP_2)^2,t\equiv -(\pP_1+\pP_3)^2)$. For future convenience, the Mellin amplitude is written as a function of $(\frac{S}{\tR},\frac{T}{\tR})$ rather than $(S,T)$.
In the flat space limit, the Mellin integral is evaluated at large $(S,T)$ along with large $\tR$ using saddle-point approximation. Deformation of the  $S,T$ contour to the steepest descent one may involve crossing some poles of the Mellin amplitude. As described in \cite{Komatsu:2020sag}, the associated residues correspond to Landau diagrams in the bulk where the internal particle propagates over distances of $O(\tR)$ while the saddle point itself corresponds to a bulk process happening over a small flat space patch. In line with our main assumption concerning the flat space limit, we will assume that the contribution of the residues is subdominant compared to that of the saddle point. We will also assume that the Mellin amplitude does not scale exponentially with $\Delta$ and hence does not control the saddle point of $S$ and $T$.

Using the Fourier transform \eqref{cft-fourier}, the momentum space CFT correlator is related to mellin amplitude as,
\begin{align}\label{M_to_P}
    \tilde{G}(\pP_{i}) &= \prod_{i=1}^{4} \Big( \int d \eW_{i} \ \delta(\eW_{i}^{2})  \Big) \ e^{i \sum_{i=1}^{4} \pP_{i} \cdot \pW_{i}} \\
    &\times (\eW_1\cdot\eW_4)^{-\Delta}(\eW_2\cdot\eW_3)^{-\Delta}\int  dS\, \int\,dT\, \Big(\frac{v}{u}\Big)^S\, v^T \,  M(\frac{S}{\tR},\frac{T}{\tR})\,\Gamma^2(S)\Gamma^2(T)\Gamma^2(\Delta-S-T)\notag
\end{align}

\subsection{At $\tR \rightarrow \infty$ limit} 

There are two kinds of integral that one needs to perform in order to get momentum space CFT correlator from mellin amplitude, as in equation \eqref{M_to_P}. One is the integral over Mellin variables $(S,T)$ and the other is the integral over positions $\pW_i$. Both of them are performed using saddle point approximation in the large $\tR$ limit.

\subsubsection*{Mellin space integral}
We scale $(S,T)\equiv (\tR \sigma,\tR \tau)$ with finite $(\sigma, \tau)$ in the large $\tR$ limit, apart from scaling $\Delta\sim m\tR$. The integral becomes
\begin{align}\label{mellin-saddle-integral}
    G(\eW_i)&=  \int dS\, \int dT\,  M(\sigma,\tau)\,\tR^{2m}\,f(\sigma,\tau) \,\,e^{\tR\,g(\sigma, \tau;\eW_i)} \notag\\
    {\rm where}\quad g(\sigma,\tau;\eW_i)&=-2m + \sigma \log \big( \frac{v}{u} \big) + \tau \log (v) +2 \sigma \log(\sigma) +2 \tau \log (\tau) \notag \\
    &+ 2(m-\sigma-\tau) \log (m-\sigma-\tau)\notag\\
    f(\sigma,\tau)&=\frac{8 \pi^{3}}{\tR^{3+d}} \frac{1}{\sigma \tau (m-\sigma-\tau)^{1+d} }.
\end{align} 
We have assumed that $M(\sigma,\tau)$ does not scale exponentially with $\tR$ and hence does not contribute to the saddle point. The integral is dominated by the saddle point of only $g(\sigma, \tau)$. This means that at leading order,
\begin{equation}\label{M_saddle_sol}
    G(\eW_i)=G_{\phi^4}(\eW_i) M(\sigma_*,\tau_*)|_{\tR\to \infty}
\end{equation}
Here $G_{\phi^4}(\eW_i)$ is the large $\tR$ limit of the contact diagram corresponding to the interaction $\phi^4$ that has Mellin amplitude $1$ and $(\sigma_*,\tau_*)$ is the saddle point of the Mellin integration.  The saddle point equations $\partial_\sigma g=\partial_\tau g=0$ are simplified to
\begin{equation}\label{mellin-saddle}
    \frac{\sigma^{2}}{(m-\sigma-\tau)^{2}} = \frac{u}{v} ,  \qquad\qquad \frac{\tau^{2}}{(m-\sigma-\tau)^{2}} = \frac{1}{v}
\end{equation}
Solving these for $(\sigma,\tau)$ we get,
\begin{equation}
    \sigma_* = \frac{m \sqrt{u}}{1+\sqrt{u}+\sqrt{v}},  \qquad  \qquad \tau_* = \frac{m }{1+\sqrt{u}+\sqrt{v}}
\end{equation}

\subsubsection*{Position space integral}
From equation \eqref{M_to_P} and \eqref{M_saddle_sol} , the CFT correlator in momentum space in large $\tR$ limit is written as,
\begin{equation}\label{pspace-limit}
    \hat{G}(\pP_{i}) \xrightarrow[]{\tR \rightarrow \infty} \prod_{i=1}^{4} \Big( \int d \eW_{i} \ \delta(\eW_{i}^{2})  \Big) \ e^{i \sum_{i=1}^{4} \pP_{i} \cdot \pW_{i}}G_{\phi^4}(\eW_i) M(\sigma_*,\tau_*).
\end{equation}
Recall,
\begin{equation}
    G_{\phi^4}(\eW_i)=\int d\eZ\,\delta(\eZ^2+\tR^2) \prod_i(\eW_i\cdot \eZ)^{-\Delta}.
\end{equation}
This is precisely the integral studied in appendix \ref{bulk-loc}. As demonstrated there, the bulk integral gets localized at $\eZ=\eC$ and the boundary integral gets localized at $\pW_i=i\pP_i/m$.

The cross-ratios at the saddle point are
\begin{equation}\label{position-saddle}
    u_*=\frac{(\pP_{1} \cdot \pP_{2} + m^2)(\pP_{3} \cdot \pP_{4} + m^2)}{(\pP_{1} \cdot \pP_{3} + m^2)(\pP_{2} \cdot \pP_{4} + m^2)}, \qquad v_*=\frac{(\pP_{1} \cdot \pP_{4} + m^2)(\pP_{2} \cdot \pP_{3} + m^2)}{(\pP_{1} \cdot \pP_{3} + m^2)(\pP_{2} \cdot \pP_{4} + m^2)}.
\end{equation}
Using equation \eqref{pspace-limit},
\begin{equation}
    \hat{G}(\pP_{i}) \xrightarrow[]{\tR \rightarrow \infty} M(\sigma_*,\tau_*)\, \hat{G}_{\phi}(\pP_{i})=M(\sigma_*,\tau_*)\,\delta\big( \sum_{i=1}^{4} \pP_{i} \big) 
\end{equation}
The second equality follows from discussion in section \ref{S-matrix-contact}. What remains now is solving for the Mellin variables $(\sigma_*,\tau_*)$ in terms of $\pP_i$ using equation \eqref{mellin-saddle} and \eqref{position-saddle} and momentum conservation $\sum_{i=1}^4 \pP_i=0$.
It is easy to check that 
\begin{equation}
    \sigma_*=(\pP_1\cdot\pP_2+m^2)/4m=(4m^2-s)/8m,\qquad \tau_*=(\pP_1\cdot\pP_3+m^2)/4m=(4m^2-t)/8m.
\end{equation}
Solves both the saddle point equations. As a result, we have
\begin{equation}
    \hat{G}(\pP_{i}) \xrightarrow[]{\tR \rightarrow \infty}  M(\frac{4m^2-s}{8m},\frac{4m^2-t}{8m})\,\delta\big( \sum_{i=1}^{4} \pP_{i} \big) 
\end{equation}
As the AdS S-matrix is defined as the part of $\hat{G}(\pP_{i})$ that is proportional to momentum conserving delta function, we get the leading order dictionary between the AdS S-matrix $S(s,t)$ and the Mellin amplitude $M(\sigma,\tau)$.
\begin{equation}
    S(s,t)=M(\frac{4m^2-s}{8m},\frac{4m^2-t}{8m}).
\end{equation}

\subsection{Sub-leading corrections}
Before computing the sub-leading corrections to the relation between the AdS S-matrix and the Mellin amplitude, let us emphasize that this relation is linear. Let $\fS$ be the linear operator that acts on the Mellin amplitude and produces the corresponding AdS S-matrix. 
\begin{align}
    S(s,t)&=\fS[M(\sigma,\tau)]\notag\\
    &=\sum_{i,j=0}^\infty\,\,\frac{1}{i!j!} \partial_\sigma^i\partial_\tau^j\,M(0,0) \,\,\fS[\sigma^i\tau^j]
\end{align}
In the second line, we have simply Taylor expanded the Mellin amplitude at $(0,0)$. As a result, the $S(s,t)$ is written as a sum of AdS S-matrices of simple contact diagrams corresponding to the Mellin amplitude $\sigma^i\tau^j$ with c-number coefficients $\partial_\sigma^i\partial_\tau^j M(0,0)/i!j!$. This equation, \emph{per se}, is not useful  in computing the relation between $S^{(0)}$ and $M$ in $1/\tR$ perturbation theory, however it suggests a useful way of looking at the saddle point approximation. The corrections to the integral \eqref{mellin-saddle-integral} are obtained by expanding the Mellin amplitude at the saddle point $(\sigma_*,\tau_*)$ rather than at $(0,0)$. This has the advantage that each derivative of $\sigma$ and $\tau$ appears with the suppression $1/\sqrt \tR$.
\begin{equation}\label{SfromM}
    S(s,t)=\sum_{i,j=0}^\infty\,\,\frac{1}{i!j!} \partial_\sigma^i\partial_\tau^j\,M(\sigma_*,\tau_*) \,\,\fS[(\sigma-\sigma_*)^i(\tau-\tau_*)^j].
\end{equation}
This neatly gives us the dictionary between the Mellin amplitude and the AdS S-matrix, we only need to compute the AdS S-matrix $\fS[(\sigma-\sigma_*)^i(\tau-\tau_*)^j]$. As emphasized, the saddle point evaluation ensures that this quantity to be of ${\cal O}(\tR^{-\frac12 (i+j)})$. As the S-matrix for any contact diagram doesn't have half-integer power of $\tR$, it turns out that in fact, $\fS[(\sigma-\sigma_*)^i(\tau-\tau_*)^j]={\cal O}(\tR^{-\lceil{\frac12 (i+j)}\rceil})$. 
Hence, it would seem that in order to compute the first non-trivial correction i.e. of ${\cal O}(1/\tR)$, we need these terms for $i+j\leq 2$. In fact, this is also  not quite correct as the S-matrix for contact diagram does not receive corrections at ${\cal O}(1/\tR)$. So the relation between the AdS S-matrix and the Mellin amplitude at ${\cal O}(1/\tR)$ ends up being a constraint, albeit novel, on the ${\cal O}(1/\tR)$ terms in the Mellin amplitude. The first non-trivial correction to the AdS S-matrix is at ${\cal O}(1/\tR^2)$. Obtaining this correction involves computing $\fS[(\sigma-\sigma_*)^i(\tau-\tau_*)^j]$ for $i+j\leq 4$. 

It is convenient to work with $\fS[\sigma^i\tau^j]$ for the same values of $(i,j)$ and take their linear combinations to obtain desired expression.
We do this by computing the AdS S-matrix and Mellin amplitude corresponding to the following types of contact diagrams
\begin{align}
    D_{i,j}\equiv&\,\, \frac12 \Big(\nabla^{a_1}\ldots\nabla^{a_i} \nabla^{b_1}\ldots\nabla^{b_j}\phi_1 \nabla_{a_1}\ldots\nabla_{a_i}\phi_2 \nabla_{b_1}\ldots\nabla_{b_j} \phi_3 \phi_4\notag\\
    &\,\,+ \nabla^{b_1}\ldots\nabla^{b_j}\nabla^{a_1}\ldots\nabla^{a_i}\phi_1 \nabla_{a_1}\ldots\nabla_{a_i}\phi_2 \nabla_{b_1}\ldots\nabla_{b_j} \phi_3 \phi_4\Big)
\end{align}
and comparing them. Using the Mellin amplitudes for the diagrams $D_{i,j}$ we construct the inverse transformation expressing the monomials $\sigma^i\tau^j$ in terms of Mellin amplitudes of these diagrams. Denoting the Mellin amplitude of $D_{i,j}$ as $M_{D_{i,j}}(\sigma,\tau)$, we write
\begin{align}\label{MtoD}
    \sigma^i\tau^j =\sum_{k,l} A^{i,j}_{k,l}\,\, M_{D_{k,l}}(\sigma,\tau).
\end{align}
We have given sample computation of $M_{D_{i,j}}$ for $i+j\leq 2$ below.
\begin{align}
    M_{D_{0,0}}&=1\notag\\
    M_{D_{1,0}}&=-\frac{2 \bar{\Delta}}{\tR} \sigma + \frac{\Delta^2}{\tR^{2}}\notag\\
    M_{D_{2,0}}&=\frac{4 \bar{\Delta}(\bar{\Delta}+1)}{\tR^{2}} \sigma^{2} - \frac{4 \bar{\Delta}( \Delta^2 + 2\Delta - \bar{\Delta})}{\tR^{3}} \sigma + \frac{\Delta^2(\Delta^2+d)}{\tR^{4}}\notag\\
    M_{D_{1,1}}&=\frac{4 \bar{\Delta}(\bar{\Delta}+1)}{\tR^{2}} \sigma \tau - \frac{2 \bar{\Delta}( \Delta^2 + 2\Delta)}{\tR^{3}} (\sigma+\tau) + \frac{\Delta^{2}(\Delta^2+ 2\bar{\Delta})}{\tR^{4}}.
\end{align}
Here $\bar \Delta=4\Delta-d$. The Mellin amplitudes $M_{D_{0,1}}, M_{D_{0,2}}$ can be computed from above using the property $M_{D_{i,j}}(\sigma,\tau)=M_{D_{j,i}}(\tau,\sigma)$.

Now in order to compute the AdS S-matrix corresponding to the Mellin amplitude, we simply evaluate the S-matrix for the diagrams $D_{i,j}$. Denoting this S-matrix as $S_{D_{i,j}}(s,t)$, we get a very simple relation
\begin{align}
    \fS[\sigma^i\tau^j]=\sum_{k,l} A^{i,j}_{k,l} S_{D_{k,l}}(s,t)
\end{align}
where $A^{i,j}_{k,l}$ are the matrix elements obtained in equation \eqref{MtoD}. 
We have given sample computation of $S_{D_{i,j}}$ for $i+j\leq 2$ below.
\begin{align}
    S_{D_{0,0}}&=1\notag\\
    S_{D_{1,0}}&=(\frac{s}{2}-m^2)\Big(1+\frac{1}{\tR^2}\frac{d}{2m^2}\Big)\notag\\
    S_{D_{2,0}}&=(\frac{s}{2}-m^2)^2\notag\\
    S_{D_{1,1}}&=(\frac{s}{2}-m^2)(\frac{t}{2}-m2)\Big(1+\frac{1}{\tR^2}\frac{d}{2m^2}\Big).
\end{align}
The AdS S-matrices $S_{D_{0,1}}, S_{D_{0,2}}$ can be computed from above using the property $S_{D_{i,j}}(s,t)=S_{D_{j,i}}(t,s)$. Although, we have not given explicit expressions for $M_{D_{i,j}}(\sigma,\tau)$ and $S_{D_{i,j}}(s,t)$ for higher values of $i,j$ - as they are cumbersome and not-so-illuminating - they can be computed in a straightforward fashion. With this, it is easy to construct the coefficients $\fS[(\sigma-\sigma_*)^i(\tau-\tau_*)^j]\equiv \fS_{i,j}$ appearing in the relation \eqref{SfromM} by taking linear combinations of  $\fS[\sigma^i\tau^j]$'s. The final result to ${\cal O}(1/\tR^2)$ is produced below. 
\begin{align}\label{sfromm}
    \fS_{0,0}&=1\notag\\
    \fS_{1,0}&=\frac{d \left(\frac{s}{m^2}+4\right)}{32 \tR}+\frac{d \left(4 (d+4) m^2+(d-8) s\right)}{128 m^3 \tR^2}\notag\\
    \fS_{2,0}&=-\frac{\left(s-4 m^2\right)^2}{128 m^3 \tR} + \frac{16 ((d-2) d+4) m^4+8 (d (d+2)-4) m^2 s+((d-10) d+4) s^2}{1024 m^4 \tR^2} \notag\\
    \fS_{1,1}&=\frac{16 m^4-4 m^2 (s+t)-s t}{128 m^3 \tR} + \frac{16 \left(d^2-4\right) m^4+4 (d (d+2)+4) m^2 (s+t)+((d-2) d+4) s t}{1024 m^4 \tR^2 }\notag\\
    \fS_{3,0}&=\frac{\left(s^2-16 m^4\right) \left(12 d m^2+(8-3 d)s\right)}{4096 m^5 \tR^2} \notag\\
    \fS_{2,1}&=\frac{4 m^2 \left(4 m^2-s\right) \left(4 d m^2+3 d s+4 s\right)-d \left(48 m^4+8 m^2 s+3 s^2\right)t}{4096 m^5 \tR^2}\notag\\
    \fS_{4,0}&=\frac{3 \left(s-4 m^2\right)^4}{16384 m^6 \tR^2}\notag\\
    \fS_{3,1}&=\frac{3 \left(s-4 m^2\right)^2 \left(4 m^2 (s+t)+s t-16 m^4\right)}{16384 m^6 \tR^2}\notag\\
    \fS_{2,2}&=\frac{768 m^8-384 m^6 (s+t)+16 m^4 \left(3 s^2+4 s t+3 t^2\right)+8 m^2 s t (s+t)+3 s^2 t^2}{16384 m^6 \tR^2} .
\end{align}
Again, the remaining $\fS_{i,j}$'s can be obtained from the property $\fS_{i,j}(s,t)=\fS_{j,i}(t,s)$.

Here we have obtained the AdS S-matrix from the Mellin amplitude to ${\cal O}(1/\tR^2)$ order. This relation can be easily inverted to obtain the Mellin amplitude from the AdS S-matrix in $1/\tR^2$ perturbation theory. We have done so below to ${\cal O}(1/\tR^2)$. As emphasized earlier, the AdS S-matrix for contact diagrams does not have $1/\tR$ corrections but the Mellin amplitude does. This gives a constraint on the ${\cal O}(1/\tR)$ Mellin amplitude. Letting 
\begin{equation}
S(s,t) = \sum_{n=0}^{\infty} \frac{1}{\tR^{2n}} S^{(n)}(s,t)\,\,, \qquad  \qquad M(\sigma,\tau) = \sum_{m=0}^{\infty} \frac{1}{\tR^{m}} M^{(m)}(\sigma,\tau),
\end{equation}
the constraint on $M^{(1)}$ takes the form,
\begin{align}
M^{(1)}(\sigma,\tau) &= \frac{d(\sigma-m)}{4m} \partial_{\sigma} M^{(0)}(\sigma,\tau) +\frac{d(\tau-m)}{4m} \partial_{\tau} M^{(0)}(\sigma,\tau) \\
\nonumber
&+\frac{\sigma^{2}}{4m} \partial_{\sigma}^{2} M^{(0)}(\sigma,\tau) +\frac{(m^2 + 2 \sigma \tau - 2 m (\sigma + \tau))}{4m}\partial_{\sigma}\partial_{\tau} M^{(0)}(\sigma,\tau)+\frac{\tau^{2}}{4m} \partial_{\tau}^{2} M^{(0)}(\sigma,\tau).
\end{align}
We have checked that the Mellin amplitude of the scalar exchange diagram indeed satisfies this constraint.
The first non-trivial correction to the Mellin amplitude is $M^{(2)}(\sigma,\tau)$. It is given in terms of $S^{(1)}(s,t)$ and $M^{(0)}(\sigma,\tau)$ in the following way.
\begin{align}
M^{(2)}(\sigma,\tau) &= S^{(1)}(4m^{2}-8m \sigma,4m^{2}-8m \tau) + \sum_{i=0}^{4} \sum_{j=0}^{4} A_{i,j}(\sigma,\tau) \ \partial_{\sigma}^{i} \partial_{\tau}^{j} M^{(0)}(\sigma,\tau) 
\end{align}
where,
\begin{align}\label{mfroms}
    A_{0,0} &= 0\notag\\
    A_{1,0} &= \frac{d ((d-4) \sigma +(1-d)m)}{8 m^2}\notag\\
    A_{2,0} &= \frac{d \left((d+2) m^2-2 (d+6) m \sigma +(d+18) \sigma ^2\right)}{32 m^2}\notag\\
    A_{1,1} &= \frac{d \left((2 d+11) m^2-2 (d+8) m (\sigma +\tau )+2 (d+10) \sigma  \tau \right)}{32 m^2}\notag\\
    A_{3,0} &= \frac{\sigma  \left(4m^{2}-3 (d+4) m \sigma +(3 d+20) \sigma ^2\right)}{48 m^2}\notag\\
    A_{2,1} &= \frac{-(d+2) m^3+m^2 ((3 d+7) \sigma +2 (d+2) \tau )-m \sigma  ((3 d+10) \sigma +4 (d+3) \tau )+3 (d+4) \sigma ^2 \tau}{16 m^2}\notag\\
    A_{4,0} &= \frac{\sigma ^4}{32 m^2}\notag\\
    A_{3,1} &= \frac{\sigma ^2 \left(m^2-2 m (\sigma +\tau )+2 \sigma  \tau \right)}{16 m^2}\notag\\
    A_{2,2} &= \frac{m^4-4 m^3 (\sigma +\tau )+4 m^2 \left(\sigma ^2+3 \sigma  \tau +\tau ^2\right)-8 m \sigma  \tau  (\sigma +\tau )+6 \sigma ^2 \tau ^2}{32 m^2}.
\end{align}
As in the previous cases, the remaining $A_{i,j}(\sigma, \tau)$'s are obtained by using $A_{i,j}(\sigma, \tau)=A_{j,i}(\tau, \sigma)$.

\section{Discussion and outlook}
In this paper, we have defined a new observable called the AdS S-matrix which in the large $\tR$ limit reduces to the familiar S-matrix in the flat space. We have developed a momentum space formalism that can be used to  compute it in ${\cal O}(1/\tR)$ perturbation theory using set of Feynman-like rules. Moreover we have established the relation between the AdS S-matrix and the more familiar CFT observable, the Mellin amplitude. Equipped with this ``momentum space'', we expect that most - if not all - of the questions about the flat space S-matrix can be investigated in AdS in $1/\tR$ perturbation theory. 
\begin{itemize}
    \item 
    Starting with \cite{Fitzpatrick:2011dm}, constraints of unitarity in AdS on CFT correlation functions have been widely explored and utilized. In the seminal paper \cite{Aharony:2016dwx} authors used crossing and unitarity to compute loop diagrams in AdS. See \cite{Meltzer:2020qbr} for an extensive list of references on loop diagram computations in AdS using unitarity methods. We expect that the AdS Cutkosky's rules should have a technically simpler formulation in conformal momentum space. This is because in the $\tR\to \infty$ limit, they simply reduce to the familiar flat space Cutkosky's rules. It would be interesting to understand these constraints and their implications on CFT data. 
    \item 
    The scattering amplitudes in flat space have been studied in great detail and many new and unexpected structures have been uncovered. See \cite{Elvang:2013cua} for a detailed review of this progress. It would be interesting to inquire if the structures such as the BCFW recursion rules \cite{Britto:2005fq}, the BCJ relation \cite{Bern:2008qj}, twistor string theory \cite{Witten:2003nn}, the amplitudehedron \cite{Arkani-Hamed:2013jha} and its more general cousin the EFT-hedron \cite{Arkani-Hamed:2020blm} admit deformations that correspond to studying the scattering in AdS in $1/\tR$ perturbation theory. The BCFW-like recursion relation has already been observed in AdS \cite{Raju:2012zr}.
    Double copy relation for CFT three point function has been studied in \cite{Farrow:2018yni, Lipstein:2019mpu}. The BCJ relation has been realized in AdS in \cite{Armstrong:2020woi}. 
    More recently, the ambitwistor string theory has been employed in AdS to compute scalar correlation functions in the dual conformal field theory \cite{Eberhardt:2020ewh} and in dS to compute cosmological correlators \cite{Gomez:2021qfd, Gomez:2021ujt}. This makes us hopeful that other structures may also have AdS counterparts and having a new perturbative parameter to study them is always helpful. 
    \item 
    Mellin space has proved to be extremely effective in computing conformal correlation functions in $1/N$ perturbation theory. Starting from the work of \cite{Rastelli:2016nze}, where the tree level four-point correlators of half-BPS operators in $\CN=4$ in the supergravity limit were computed in Mellin space, the technique has been generalized to compute AdS loops in supergravity \cite{Alday:2017xua} as well as string theory \cite{Alday:2018pdi}. It would be interesting to see if the conformal momentum space defined here would be useful to study these correlators. 
    \item 
    The Mellin space is usually quite convenient for doing Witten diagram computations.  However, calculating the Mellin amplitude for Witten diagrams containing spinning  particles is technically somewhat cumbersome. This is because the polarizations  couple to momenta and not kinematical invariants. The Mellin space only offers variables of the latter type. We expect that incorporating spin should be easier with the ``momentum space'' developed here as the polarizations simply couple to $\pP$'s.
\end{itemize}
At this point, let us also highlight a conceptual issue that arises in relating the boundary correlator to the AdS S-matrix already at leading order. As described in \cite{Komatsu:2020sag} the physical scattering regime in the Mandelstam space is related to the correlator on the so called ``Regge-sheet'' of the cross-ratios. However, as pointed out in \cite{Heemskerk:2009pn, Maldacena:2015iua}, a physical scattering experiment can be set up in the bulk only when the boundary points are in a configuration that corresponds to a different sheet in cross-ratio space, termed the ``scattering-sheet'' in \cite{Chandorkar:2021viw}. Indeed, the bulk-point singularity corresponding to massless scattering is present on the scattering sheet and not on the Regge sheet\footnote{We thank Shiraz Minwalla for extensive discussion on this issue.}. As explained in \cite{Hijano:2019qmi} and also in \cite{Komatsu:2020sag}, the Regge sheet correlator for operators corresponding to massive particles in the flat space limit, can be understood as happening in a partially Wick rotated Lorentzian AdS geometry \emph{a.k.a} AdS with caps. This is the same setup studied in \cite{Skenderis:2008dg,Skenderis:2008dh}. We feel that this dichotomy between flat space limit of massive and massless particle Witten diagrams needs to be explored further.

Let us also comment on the on-shell states for massless particles. For that, we need $\pP_i^2=0$. The Fourier transform \eqref{cft-fourier} does not naively seem to allow for this condition. It would be useful to think about this issue further if we would like to generalize the technology introduced in the paper to massless particles. It could help in computing the sub-leading terms in the bulk-point singularity expansion.

Lastly, let us note that although we have defined AdS S-matrix for a non-perturbative theory in the bulk using CFT correlation function, the investigation in this paper focuses completely on perturbative theories in the bulk. It would interesting to compute the non-perturbative S-matrices in $1/\Delta$ expansion from a known CFT correlation functions, such as those of rational conformal field theories in $2d$ and analyze their properties.

\section*{Acknowledgements}
We would like to thank Sayali Bhatkar, Sachin Jain and the TIFR string theory group, in particular Gautam Mandal, Shiraz Minwalla, Onkar Parrikar, Sandip Trivedi for useful discussions. We would like to thank Shraiyance Jain for collaboration during the initial stage of the work.
This work is supported by the Infosys Endowment for the study of the Quantum Structure of Spacetime and by the SERB Ramanujan fellowship.  We acknowledge support of the Department of Atomic Energy, Government of India, under Project Identification No. RTI 4002. We would also like to acknowledge our debt to the people of India for their steady support to the study of the basic sciences.

\appendix

\section{Exact propagators}\label{exact-prop}
To get some intuition about the Fourier transform \eqref{ads-fourier} and \eqref{cft-fourier}, let us do a very simple toy integral. Let us work in the embedding space with signature $(+,\ldots,+,-)$.
\begin{align}\label{toy}
    I(\eT)=\int_{+} \, d\eX \,\delta(\eX\cdot \eX+\tR^2)\, e^{\eT\cdot \eX}
\end{align}
This integral is useful in carrying out Fourier transforms with respect to the points in the bulk as well, as in equation \eqref{ads-fourier}. In the limit $\tR\to 0$ this becomes be the boundary integral. The components of these embedding space vectors are $\eT=(\eT^+,\eT^-,t^\mu)$ and the same for $\eX$. The dot product is 
\begin{align}
    \eT\cdot \eX=t_\mu x^\mu-\frac12(\eT^+\eX^-+\eT^-\eX^+).
\end{align}
The subscript $+$ on the integration sign denotes that the integral is supported only over $\eX^+\geq 0$. This means it is supported only over the upper hyperboloid (in the case $\tR^2>0$) or on the positive null cone (when we set $\tR=0$).

We first first solve for $\eX^-$ using the fact that the integral is supported only over $\eX^2=\tR^2$.
\begin{align}
    I(\eT)=\int_0^\infty d\eX^+\,\int  dx^\mu \, \frac{1}{\eX^+}e^{t_\mu x^\mu-\frac12(\eT^-\eX^+ +(x^2+\tR^2) \frac{\eT^+}{\eX^+})}.
\end{align}
Integral over $x^\mu$ is Gaussian and is easily done.
\begin{align}
    I(\eT)=\int_0^\infty d\eX^+\,\frac{1}{\eX^+}\Big(\frac{2 \pi \eX^+}{\eT^+}\Big)^{\frac{d}{2}}e^{\frac12 \Big(\eT^2 \frac{\eX^+}{ \eT^+}- \tR^2\frac{ \eT^+}{\eX^+}\Big)}
\end{align}
The $\eX^+$ is identified as the integral representation of the modified Bessel function of the second kind. Finally we have,
\begin{align}
    I(\eT)=2(2\pi)^{\frac{d}{2}}\tR^{\frac{d}{2}}(-\eT^2)^{-\frac{d}{4}} K_{-\frac{d}{2}}(\tR\sqrt{-\eT^2}).
\end{align} 
In the limit $\tR\to 0$, the Bessel function simplifies to give,
\begin{align}
    I_{0}(\eT)\equiv I_{\tR\to 0}(\eT)=(2\pi)^{\frac{d}{2}}\Gamma(d/2) (-\eT^2)^{-\frac{d}{2}}.
\end{align}
Upto multiplicative constant, this final result is expected because in the $\tR\to 0$ limit, the integral \eqref{toy} has to be a function only of $\eT^2$ and it homogenous in $\eT$ with degree $-d$.

\subsection{Boundary two-point function}
In this section, we compute the boundary two point function in the $\eP$-space. In position space, the boundary two point function takes a very simple form.
\begin{equation}
G(\eW_1,\eW_2) = \frac{1}{(- \eW_1 \cdot \eW_2)^{\Delta}}
\end{equation} 
where, $\eW_i$  are boundary points i.e. they obey $\eW_i \cdot \eW_i=0$. The $\eP$ space two-point function is computed by the Fourier transform \eqref{cft-fourier}.
\begin{align}
    \tilde G(\eP_1,\eP_2) =  \int_+\, d\eW_i\, \delta(\eW_i^2)\,  \ e^{i \sum_{i=1}^2  \eP_i \cdot \eW_i} \frac{1}{(- \eW_1 \cdot \eW_2)^{\Delta}}
\end{align}
Using  the Schwinger representation for the position space propagator, we rewrite the integral as 
\begin{equation}
    \tilde G(\eP_1,\eP_2) = \int_+\, d\eW_i \, \delta(\eW_i^2)\,  \ e^{i \sum_{i=1}^2  \eP_i \cdot \eW_i} \frac{1}{\Gamma(\Delta)}\int_0^\infty \frac{dt}{t}t^{\Delta} e^{t(\eW_1\cdot \eW_2)}. 
\end{equation}
This integral can be done using the results above. Performing integral over $\eW_{1}$ we get,
\begin{equation}
    \tilde G(\eP_1,\eP_2) =  \frac{1}{\Gamma(\Delta)}\int_0^\infty \frac{dt}{t}t^{\Delta} \int_+\, d\eW_2 \, \delta(\eW_2^2)\,  \ e^{i \eP_2 \cdot \eW_2} \ (2\pi)^{\frac{d}{2}} \Gamma(d/2)(-|i \eP_1+t \eW_2|^{2})^{-\frac{d}{2}}. 
\end{equation}
To perform the integral over $\eW_{2}$, we first exponentiate the factor $(-|i \eP_1+t \eW_2|^{2})$ by introducing another Schwinger parameter (say $s$), and then use the above results to get,
\begin{equation}
    \tilde G(\eP_1,\eP_2) =  \frac{1}{\Gamma(\Delta)}\int_0^\infty \frac{dt}{t}t^{\Delta} \ \frac{1}{\Gamma(d/2)}\int_0^\infty \frac{ds}{s}s^{d/2} \ ((2\pi)^{\frac{d}{2}} \Gamma(d/2))^{2} \  e^{- s   |\eP_1|^2} \ (-|i \eP_2 + 2i s t \eP_1|^{2})^{-\frac{d}{2}}. 
\end{equation}

Doing the change of variables from $t$ to $t'/s$, the integrals over $s$ and $t'$ decouple. Performing $s$ integral we get,
\begin{equation}
    \tilde G(\eP_1,\eP_2) =  \frac{(2\pi)^{d} \Gamma(d/2)}{\Gamma(\Delta)}\Gamma(d/2-\Delta) \ |\eP_{1}|^{2 \Delta-d} \ \int_0^\infty \frac{dt'}{t'}(t')^{\Delta}   \ (-|i \eP_2 + 2i t' \eP_1|^{2})^{-\frac{d}{2}}. 
\end{equation}
The $t'$ integral can be performed by first exponentiating $(-| i \eP_2 + 2i t' \eP_1|^2 )^{-\frac{d}{2}}$ by introducing another Schwinger parameter $r$, then integrating out $t'$, followed by integration over $r$. This gives,
\begin{align}
\nonumber
\hat G(\eP_1,\eP_2) &=  \frac{(2\pi)^{d} \Gamma(d/2-\Delta)}{2^{\Delta+1}\Gamma(\Delta)} \  \Big( \Gamma \Big( \frac{d-\Delta}{2} \Big) \Gamma \Big( \frac{\Delta}{2} \Big) \  _{2}F_{1}\Big[\frac{d-\Delta}{2}, \frac{\Delta}{2}, \frac{1}{2} , \frac{(\eP_1 \cdot \eP_2)^2}{|\eP_1|^2|\eP_2|^2} \Big] \\
&  -2 \ \Gamma \Big( \frac{d-\Delta+1}{2} \Big) \Gamma \Big( \frac{\Delta+1}{2} \Big) \ \frac{(\eP_1 \cdot \eP_2)}{|\eP_1| \ |\eP_2|} \ _{2}F_{1}\Big[\frac{d-\Delta+1}{2}, \frac{\Delta+1}{2}, \frac{3}{2} , \frac{(\eP_1 \cdot \eP_2)^2}{|\eP_1|^2|\eP_2|^2} \Big] \Big)
\end{align}
We get the  exact expression for  $\hat G(\eP_1,\eP_2)$ that is given in \eqref{exact-2} where the constants $\alpha$ and $\beta$ are,
\begin{align}
    \alpha= \frac{(2\pi)^{d} \Gamma(d/2-\Delta)}{2^{\Delta+1}\Gamma(\Delta)} \Gamma \Big( \frac{d-\Delta}{2} \Big) \Gamma \Big( \frac{\Delta}{2} \Big),\qquad \beta = - \frac{(2\pi)^{d} \Gamma(d/2-\Delta)}{2^{\Delta}\Gamma(\Delta)} \Gamma \Big( \frac{d-\Delta+1}{2} \Big) \Gamma \Big( \frac{\Delta+1}{2} \Big)
\end{align}

\section{Bulk localization in contact Witten diagrams}\label{bulk-loc}
In this appendix we show that the integral over the bulk point in an $n$-point contact Witten diagram localizes at the particular point $\eC=(0,\ldots, 0, \pm\tR)$ in the large $\tR$ limit, if we fix the gauge $\eP^{(i)}_{d+1}=0$ for the ``momenta'' variables of all the insertions. For simplicity we will deal with the Contact diagram $\phi^n$ although it is straightforward to extend our treatment to contact diagrams involving derivatives. For generality we will take the conformal dimensions of $i$-th operator to be $\Delta_i$.
In what follows $(\eQ_i,\eP_i)$ are the positions and conformal momenta of the boundary points and $\eZ$ is the bulk point to be integrated over.
The $n$-point contact Witten diagram in position space is
\begin{align}
D(\eQ_{i}) &= \int d \eZ\, \delta(\eZ^2+\tR^2) \, \prod_{i=1}^{n} G(\eQ_i,\eZ)
\end{align}
In $\eP$ space we get the diagram by Fourier transforming as defined in equation \eqref{cft-fourier-2}. Note that it is convenient to use the homogeneous version of the Fourier transform over the usual one i.e. in equation \eqref{cft-fourier}.
\begin{align}
\tilde D(\eP_{i}) &= \int\, \prod_{i=1}^{n}d \eQ_{j}\,  \delta(\eQ_j^2) \, (\eP_{j} \cdot \eQ_{j})^{\Delta_{j}-d} D(\eQ_i) \\
&= \int d \eZ \delta(\eZ^2+\tR^2) \ \prod_{i=1}^{n} \Big[ \int d \eQ_{i} \delta(\eQ_i^2)\ (\eP_{i} \cdot \eQ_{i})^{\Delta_{i}-d} G (\eQ_i,\eZ) \Big]
\end{align}
In the first step we have dropped the numerical constant as it does not affect the saddle point computation. The term in square bracket in second equality is the bulk to boundary propagator in momentum space $ \tilde{G} (\eP_i,\eZ)$. 
\begin{align}
    \tilde{G} (\eP_i,\eZ)=\int d \eQ_{i}\,\, d\lambda_i e^{\frac12 \lambda_i \eQ_i^2}\,\, (\eP_{i} \cdot \eQ_{i})^{\Delta_{i}-d} (\eQ_i\cdot\eZ)^{-\Delta}.
\end{align}
In the large $\tR$ limit, we take $\Delta_i$ to be large. Using the fact that $\lambda_i$ is also large, this integral can be done by saddle point computation. 
The relevant function to extremize is then,
\begin{equation}
g_{i}(\eQ_{i}, \lambda_{i}) = \Delta_{i} \log \Big( \frac{\eP_{i} \cdot \eQ_{i}}{\eQ_i \cdot \eZ} \Big) + \frac{\lambda_{i}}{2} (\eQ_{i} \cdot \eQ_{i})
\end{equation} 
Saddle point equations turn out to be,
\begin{align}\label{SPG}
\frac{\partial}{\partial \lambda_{i}} g_{i}(\eQ_{i}, \lambda_{i}) \big|_{\eQ_{i}=\eQ^{*}_{i}, \lambda_{i}= \lambda^{*}_{i}} &= \eQ^{*}_{i} \cdot \eQ^{*}_{i} =0 \\
\Big(\frac{\partial}{\partial \eQ_{i}} \Big)^{A} g_{i}(\eQ_{i}, \lambda_{i}) \big|_{\eQ_{i}=\eQ^{*}_{i} , \lambda_{i}= \lambda^{*}_{i} } &= \Delta_{i} \frac{\eP_{i}^{A}}{\eP_{i} \cdot \eQ^{*}_{i}} - \Delta_{i} \frac{\eZ^{A}}{\eQ^{*}_{i} \cdot \eZ} + \lambda^{*}_{i} \eQ_{i}^{* A} = 0
\end{align}
Thus, the bulk to boundary propagator in $\eP$-space becomes,
\begin{equation}\label{bulk-boundary-saddle}
\tilde{G}(\eP_i,\eZ) \sim \Big( \frac{\eP_{i} \cdot \eQ^{*}_{i}}{\eQ^{*}_i \cdot \eZ} \Big)^{\Delta_{i}}
\end{equation}
In this step again, we have dropped the non-exponential in $\tR$ factors that do not affect the saddle point locus of the bulk point $\eZ$. Using the following tricks we solve for $\theta_i\equiv (\eP_i\cdot \eQ_i^*/\eZ\cdot \eQ_i^*)$.
\begin{align}
\eZ \cdot \frac{\partial}{\partial \eQ_{i}} g_{i}(\eQ_{i}, \lambda_{i}) \big|_{\eQ_{i}=\eQ^{*}_{i} , \lambda_{i}= \lambda^{*}_{i} } =0 & \ \Rightarrow \ \lambda^{*}_{i} = \frac{\Delta_{i}}{\eQ^{*}_{i} \cdot \eZ} \Big( \frac{ |\eZ|^{2} }{\eQ^{*}_i \cdot \eZ} - \frac{\eP_{i} \cdot \eZ}{\eP_i \cdot \eQ^{*}_{i}} \Big) \\
\eP_{i} \cdot \frac{\partial}{\partial \eQ_{i}} g_{i}(\eQ_{i}, \lambda_{i}) \big|_{\eQ_{i}=\eQ^{*}_{i} , \lambda_{i}= \lambda^{*}_{i} } =0 & \ \Rightarrow \ \lambda^{*}_{i} = \frac{\Delta_{i}}{ \eP_{i} \cdot \eQ^{*}_{i} } \Big( \frac{ \eP_{i} \cdot \eZ  }{\eQ^{*}_i \cdot \eZ} - \frac{|\eP_{i}|^{2}}{\eP_i \cdot \eQ^{*}_{i}} \Big)
\end{align}
Equating the above two expressions of $\lambda^{*}_{i}$ one gets quadratic equation in $\theta_{i}$ with solution as,
\begin{equation}\label{theta-sol}
\theta_i^{2} |\eZ|^2 -2 \theta_i \eP_{i} \cdot \eZ + |\eP_{i}|^2 =0 \quad \Rightarrow \quad \theta_i = \frac{\eP_{i} \cdot \eZ \pm \sqrt{(\eP_{i} \cdot \eZ)^2 - |\eP_{i}|^2 |\eZ|^2} }{|\eZ|^2}.
\end{equation}

Now we move to the bulk integral. The Witten diagram is
\begin{align}
    \tilde D(\eP_i)=\int d\eZ \, \, \int\, d\rho e^{\frac 12 \rho (\eZ^2+\tR^2)}\,\,\prod_{i=1}^n\,\, \tilde G(\eP_i\cdot \eZ). 
\end{align}
Using the  expression \eqref{bulk-boundary-saddle} for the bulk to boundary propagator in the large $\tR$ limit, along with equation \eqref{theta-sol},
\begin{align}
    \tilde D(\eP_i)=\int d\eZ \, \, \int\, d\rho e^{\frac 12 \rho (\eZ^2+\tR^2)}\,\,\prod_{i=1}^n\,\, \Big( \frac{\eP_{i} \cdot \eZ \pm \sqrt{(\eP_{i} \cdot \eZ)^2 - |\eP_{i}|^2 |\eZ|^2} }{|\eZ|^2} \Big)^{\Delta_{i}}. 
\end{align}
To compute the saddle point, now we need to extremize the function, 
\begin{equation}
f(\eZ,\rho) = \sum_{i=1}^{n} \Delta_{i} \log \Big( \frac{\eP_{i} \cdot \eZ \pm \sqrt{(\eP_{i} \cdot \eZ)^2 - |\eP_{i}|^2 |\eZ|^2} }{|\eZ|^2} \Big) + \frac{\rho}{2}(\eZ \cdot \eZ + \tR^2).
\end{equation}
The saddle point equations are,
\begin{align}
\frac{\partial}{\partial \rho} f(\eZ, \rho) \big|_{\eZ=\eZ^{*}, \rho= \rho^{*}} &= \eZ^{*} \cdot \eZ^{*} +\tR^2 =0 \\
\eZ \cdot \frac{\partial}{\partial \eZ}  f(\eZ, \rho) \big|_{\eZ=\eZ^{*}, \rho= \rho^{*}} &= \rho^{*} \eZ^{*} \cdot \eZ^{*} - \sum_{i=1}^{n} \Delta_{i} =0
\end{align}
Using these, one can eliminate $\rho^{*}$ to get,
\begin{equation}\label{SPF}
\Big(\frac{\partial}{\partial \eZ} \Big)^{A} f(\eZ, \rho) \big|_{\eZ=\eZ^{*}, \rho= \frac{\sum_{i=1}^{n} \Delta_{i}}{\eZ^{*} \cdot \eZ^{*}}, \eZ^{*} \cdot \eZ^{*} = -\tR^{2}} = \sum_{i=1}^{n} \frac{\Delta_{i}}{\sqrt{(\eP_{i} \cdot \eZ^{*})^2 + |\eP_{i}|^2 \tR^2}} \Big(  \eP_{i}^{A} + \frac{\eP_{i} \cdot \eZ^{*}}{\tR^2} \eZ^{* A} \Big) =0
\end{equation}
where, we have absorbed the $\pm$ sign in $\eP_{i}$. Let us define,
\begin{equation}\label{k}
\mathbf{k}_{j}^{A} = i \frac{\Delta_{j}}{\sqrt{(\eP_{j} \cdot \eZ^{*})^2 + |\eP_{j}|^2 \tR^2}} \Big(  \eP_{j}^{A} + \frac{\eP_{j} \cdot \eZ^{*}}{\tR^2} \eZ^{* A} \Big)
\end{equation}
These $\mathbf{k}_{j}$s are on-shell and are tangent vectors to AdS, i.e,
\begin{equation}
\mathbf{k}_{j} \cdot \mathbf{k}_{j} = - \frac{\Delta_{j}^{2}}{\tR^{2}} = - m_{j}^{2} \qquad ; \qquad \mathbf{k}_{j} \cdot \eZ^{*} =0
\end{equation}
Thus, the $\mathbf{k}_{j}$s can be thought of as local momentum at the interaction region (neighbourhood near point $\eZ^{*}$). Moreover, the saddle point equation \eqref{SPF} reduces to momentum conserving condition in $\mathbf{k}_{j}$ i.e.,
\begin{equation}
\sum_{j=1}^{n} \mathbf{k}_{j}^{A}=0.
\end{equation}

For a given external conformal momenta $\eP_{i}$'s one can find the interaction point $\eZ^{*}$ from saddle point equation \eqref{SPF} and hence the local bulk on-shell momenta entering interaction region at point $\eZ^{*}$ from \eqref{k}. But let us find in the $\eP_{i}$ space where the Witten diagram $\tilde D(\eP_i)$ maximizes. To do so we need to find the saddle point of $\tilde D(\eP_i)$ wrt $\eP_{i}$. This can be done leg by leg. In what follows, it is convenient to pick the gauge $\eP_i^{d+1}=0$.  With this, $\eP_{i} \cdot \eZ = \pP_{i} \cdot \pZ$ and  $\eP_{i} \cdot \eP_{i} = \pP_{i} \cdot \pP_{i}$. The function to extremize here is,
\begin{equation}
r(\pP_{i},\gamma) = \Delta_{i} \log \Big( \pP_{i} \cdot \pZ  \pm \sqrt{(\pP_{i} \cdot \pZ )^{2} + (\pP_{i} \cdot \pP_{i}) \tR^{2} } \ \Big) + \frac{\gamma}{2} (\pP_{i} \cdot \pP_{i} + m_{i}^{2})
\end{equation} 
where, $\gamma$ is the lagrange multiplier whose extremization restricts $\pP_{i}$ to be on-shell i.e. $\partial_{\gamma}r(\pP_{i},\gamma) = \pP^{*}_{i} \cdot \pP^{*}_{i} + m_{i}^{2}=0 $. It is easy to solve for $\gamma^*$ and $\pP_i^*\cdot \pZ$. Putting these solutions back in the vectorial saddle point equation of $\pP_i$, we get
\begin{equation}\label{SPR}
\pP_{i}^{* a} \sqrt{(\eZ^{d+1})^{2} - \tR^{2}} -  \sqrt{-m_{i}^{2}} \ \pZ^{a}=0 \quad \forall \ i
\end{equation}
This equation when solved together with equation \eqref{SPF} by picking the $d+1$ component of the vector, we get
\begin{equation}
\eZ^{* d+1} = \pm \tR, \quad  \qquad \pZ^{* a} =0 \qquad {\rm i.e.} \qquad \eZ^*=\eC.
\end{equation} 
Now all the equations simplify. We get
\begin{equation}
\mathbf{k}_{i}^{a} = \pP_{i}^{* a} \qquad ; \qquad \mathbf{k}_{j}^{d+1} = 0 
\end{equation}
and moreover the momentum conservation in $\mathbf{k}_{i}^{a}$ becomes momentum conservation in $ \pP_{i}^{* a}$ 
\begin{equation}
\sum_{i=1}^{n} \pP_{i}^{* a}=0. 
\end{equation}

To summarize, we have shown that  the momentum space Witten contact diagram maximizes for the external momenta $\pP_{i}$ to be on total momentum conserving locus. This means that in large $\tR$ limit $\tilde D(\pP_{i})$ is exponentially large in $\tR$ when $\sum_{i=1}^{n} \pP_{i}^{a} $ is equal to zero compared to when it is not. This behaves like as if $\tilde D(\pP_{i})$ is proportional to Dirac delta function $\delta(\sum_{i=1}^{n} \pP_{i}^{a})$. And, for such momenta configuration the bulk scattering process in large $\tR$ limit localizes at $\eZ^{A} = \{ 0,\pm \tR \}$. 

Note that here we have chosen the guage $\eP^{d+1}_i = 0$. However, the solution for $\eZ^*$ is the same even if we choose the gauge $\eP^{d+1}_i = \kappa_i$ for some fixed $\kappa_i$.

\bibliography{AdS-momentum}
\end{document}